\documentclass[twocolumn,english,prb,aps,english]{revtex4-1}
\usepackage[T1]{fontenc}
\usepackage[latin9]{inputenc}
\usepackage{float}
\usepackage{amsmath}
\usepackage{amssymb}
\usepackage{graphicx}
\usepackage{setspace}
\usepackage{esint}

\makeatletter

\@ifundefined{textcolor}{}
{%
 \definecolor{BLACK}{gray}{0}
 \definecolor{WHITE}{gray}{1}
 \definecolor{RED}{rgb}{1,0,0}
 \definecolor{GREEN}{rgb}{0,1,0}
 \definecolor{BLUE}{rgb}{0,0,1}
 \definecolor{CYAN}{cmyk}{1,0,0,0}
 \definecolor{MAGENTA}{cmyk}{0,1,0,0}
 \definecolor{YELLOW}{cmyk}{0,0,1,0}
}

\@ifundefined{textcolor}{}{%
 \definecolor{BLACK}{gray}{0}
 \definecolor{WHITE}{gray}{1}
 \definecolor{RED}{rgb}{1,0,0}
 \definecolor{GREEN}{rgb}{0,1,0}
 \definecolor{BLUE}{rgb}{0,0,1}
 \definecolor{CYAN}{cmyk}{1,0,0,0}
 \definecolor{MAGENTA}{cmyk}{0,1,0,0}
 \definecolor{YELLOW}{cmyk}{0,0,1,0}
}

\usepackage{dcolumn}
\usepackage{bm}
\usepackage{enumerate}

\usepackage{babel}

\usepackage{babel}

\makeatother

\usepackage{babel}
\begin{document}

\title{Time-dependent Stochastic Bethe-Salpeter Approach}

\author{Eran Rabani}

\affiliation{Department of Chemistry, University of California and Lawrence Berkeley
National Laboratory, Berkeley, California 94720, USA}

\author{Roi Baer}

\affiliation{Fritz Haber Center for Molecular Dynamics, Institute of Chemistry,
The Hebrew University of Jerusalem, Jerusalem 91904, Israel}

\author{Daniel Neuhauser}

\affiliation{Department of Chemistry and Biochemistry, University of California,
Los Angeles, CA-90095 USA}
\begin{abstract}
A time-dependent formulation for electron-hole excitations in extended
finite systems, based on the Bethe-Salpeter equation (BSE), is developed
using a stochastic wave function approach. The time-dependent formulation
builds on the connection between time-dependent Hartree-Fock (TDHF)
theory and configuration-interaction with single substitution (CIS)
method. This results in a time-dependent Schrödinger-like equation
for the quasiparticle orbital dynamics based on an effective Hamiltonian
containing direct Hartree \textit{\emph{and }}\textit{screened} exchange
terms, where screening is described within the Random Phase Approximation
(RPA). To solve for the optical absorption spectrum, we develop a
stochastic formulation in which the quasiparticle orbitals are replaced
by stochastic orbitals to evaluate the direct and exchange terms in
the Hamiltonian as well as the RPA screening. This leads to an overall
quadratic scaling, a significant improvement over the equivalent symplectic
eigenvalue representation of the BSE. Application of the time-dependent
stochastic BSE (TD\textit{s}BSE) approach to silicon and CdSe nanocrystals
up to size of $\approx3000$ electrons is presented and discussed.
\end{abstract}
\maketitle

\section{Introduction\label{sec:intro}}

Understanding electron-hole excitations in large molecular systems and
nanostructures is essential for developing novel optical and
electronic devices.\cite{Coe2002a,Tessler2002,Gur2005,Talapin2005}
This is due, for example, to the exponential sensitivity of the
photo-current characteristics to the excitonic energy levels and the
sensitivity of the device performance to the optical oscillator
strength. It becomes, therefore, a necessity to develop accurate
theoretical tools to describe the excitonic level alignment and the
absorption spectrum, with computational complexity that is scalable to
systems of experimental relevance (thousands of atoms and more).

\begin{singlespace}
There is no doubt that time-dependent density functional theory
(TDDFT)~\cite{Runge1984} has revolutionized the field of electronic
spectroscopy of small molecular
entities.\cite{Leeuwen2001,Onida2002,Maitra2002a,Marques2004,Burke2005a,Botti2007,Jacquemin2009,Casida2009,Adamo2013}
TDDFT provides access to excited state energies, geometries, and other
properties of small molecules with a relatively moderate computational
cost, similar to configuration interaction with single substitutions
(CIS) in the linear response frequency-domain
formulation~\cite{Stratmann1998} ($O\left(N^{4}\right)$, where $N$ is
the number of electrons), or even better using a real-time
implementation~\cite{Yabana1996,Bertsch2000,Baer2004b}
($O\left(N^{2}\right)$). In principle TDDFT is exact but in practice
approximations have to be introduced. The most common is the so-called
time-dependent Kohn-Sham (TDKS) method within the adiabatic
approximation, which has been applied to numerous challenging
problems~\cite{Bauernschmitt1996,Bauernschmitt1998,Fabian2001,Vasiliev2002,Shao2003,Troparevsky2003,Maitra2005,Andzelm2007,Govind2009,Peach2009,Kuritz2011,Peach2011,Srebro2011a,Richard2011,Chantzis2013,Bauernschmitt1997,Chelikowsky2003,Gavnholt2009,Hirata1999,Hirata1999c,Jacquemin2008,Stein2009,Stein2009a,Phillips2011,Ou2014}
with great success. However, TDKS often
fails,\cite{Parac2003,Grimme2003,Dreuw2004,Maitra2004,Hieringer2006,Levine2006,Lopata2011,Kowalczyk2011,Isborn2013}
particularly for charge-transfer excited states, multiple excitations,
and avoided crossings. In the present context, perhaps the most
significant failure of TDKS is in the description of low-lying
excitonic states in
bulk.\cite{Albrecht1998,Rohlfing2000,Sottile2007,Ramos2008,Rocca2012a}

An alternative to TDDFT, which has mainly been applied to condensed
periodic structures, is based on many-body perturbation theory (MBPT).
The most common flavors are the GW approximation~\cite{Hedin1965}
to describe quasiparticle excitations ($G$ indicates the single-particle
Green function and $W$ the screened Coulomb interaction) and the
Bethe-Salpeter equation (BSE)~\cite{Salpeter1951} to describe electron-hole
excitations. Both approaches offer a reliable solution to quasiparticle~\cite{Hybertsen1985,Hybertsen1986,Steinbeck1999,Rieger1999,Rinke2005,Neaton2006,Tiago2006,Friedrich2006,Gruning2006,Shishkin2007,Huang2008,Rostgaard2010,Tamblyn2011,Liao2011,Refaely-Abramson2011,Marom2012,Isseroff2012,Refaely-Abramson2012,Kronik2012}
and optical~\cite{Albrecht1998,Benedict1998,Rohlfing2000,Benedict2003,Spataru2004,Tiago2006,Sai2008,Fuchs2008,Ramos2008,Palummo2009,Schimka2010,Rocca2010,Blase2011,Rocca2012a,Faber2012,Faber2014}
excitations, even for situations where TDKS often fails, for example
in periodic systems~\cite{Albrecht1998,Benedict1998,Rohlfing2000,Sottile2007,Ramos2008,Rocca2012a}
or for charge-transfer excitations in molecules.\cite{Rocca2010}
However, the computational cost of the MBPT methods is considerably
more demanding than for TDKS, because conventional techniques require
the explicit calculation of a large number of occupied and virtual
electronic states and the evaluation of a large number of screened
exchange integrals between valence and conduction states. This leads
to a typical scaling of $O\left(N^{6}\right)$ and limits the practical
applications of the BSE to small molecules or to periodic systems
with small unit cells. 

Significant progress has been made by combining ideas proposed in the
context of TDDFT~\cite{Walker2006,Rocca2008} and techniques used to
represent the dielectric function~\cite{Wilson2008} based on density
functional perturbation theory.\cite{Baroni2001} This leads to an
approach that explicitly requires only the occupied orbitals (and not
the virtual states) and thus scales as $O\left(N^{2}\times
N_{k}^{2}\times N_{g}\right)$,\cite{Rocca2012a} where $N_{k}$ the
number of points in the Brillouin zone and $N_{g}$ the size of the
basis. Even with this more moderate scaling, performing a
Bethe-Salpeter (BS) calculation for large systems with several
thousands of electrons is still prohibitive.

Recently, we have proposed an alternative formulation for a class of
electronic structure methods ranging from the density functional
theory (DFT),\cite{Baer2013,Neuhauser2014} M{\o}ller-Plesset second
order perturbation theory (MP2),\cite{Neuhauser2013,Ge2013} the random
phase approximation (RPA) to the correlation
energy,\cite{Neuhauser2013a} and even for multiexciton generation
(MEG).\cite{Baer2012a} But perhaps the most impressive formulations
are that for calculating the quasiparticle energy within the GW
many-body perturbation correction to DFT~\cite{Neuhauser2014a} and for
a stochastic TDDFT.\cite{Gao2014TDsDFT} The basic idea behind our
formulation is that the occupied and virtual orbitals of the Kohn-Sham
(KS) Hamiltonian are replaced by stochastic orbitals and the density
and observables of interest are determined from an average of
stochastic replicas in a trace formula. This facilitates
\textquoteleft \textquoteleft self-averaging\textquoteright
\textquoteright{} which leads to the first ever report of sublinear
scaling DFT electronic structure method (for the total energy per
electron) and nearly linear scaling GW approach, breaking the
theoretical scaling limit for GW as well as circumventing the need for
energy cutoff approximations.

In this paper we develop an efficient approach for calculating electron-hole
excitations (rather than charge excitations) based on the BSE, making
it a practical and accessible computational tool for very large molecules
and nanostructures. The BSE is often formulated in the frequency domain
and thus requires the calculation of \textit{screened} exchange integrals
between occupied and virtual states. Instead, we introduce concepts
based on stochastic orbitals and reformulate the BSE in the time-domain
as means of reducing CPU time and memory. The real-time formulation
of the BSE delivers the response function (and thus the optical excitation
spectrum) without requiring full resolution of the excitation energies,
thereby reducing dramatically the computational cost. This is demonstrated
for well-studied systems of silicon and CdSe nanocrystals, covering
the size range of $N\approx100-3000$ electrons. Within this range,
we show that the approach scales quadratically ($O\left(N^{2}\right)$)
with system size. 
\end{singlespace}

\section{Theory \label{sec:TD-sDFT}}

In this section we review the symplectic eigenvalue formulation of
the BSE and then build on the connections between configuration interaction
with single substitution (CIS) and time-dependent Hartree-Fock (TDHF)
to formulate a time-dependent wave-equation for the BSE.

\subsection{Symplectic Eigenvalue Bethe-Salpeter Equation}

\begin{singlespace}
Within linear response, one can show that the BSE is equivalent to
solving the symplectic eigenvalue problem~\cite{Casida1995,Casida1996,Furche2001}

\begin{equation}
{\cal L}\left(\begin{array}{c}
X\\
Y
\end{array}\right)=\hbar\omega\left(\begin{array}{cc}
1 & 0\\
0 & -1
\end{array}\right)\left(\begin{array}{c}
X\\
Y
\end{array}\right)\label{eq:casida}
\end{equation}
where
\end{singlespace}

\begin{equation}
{\cal L}=\begin{pmatrix}A & B\\
-B & -A
\end{pmatrix}\label{eq:L}
\end{equation}
with
\begin{align}
A & =D+2K^{X}+K^{D}\nonumber \\
B & =2K^{X}+K^{D}.\label{eq:A+B}
\end{align}
The diagonal ($D$), exchange ($K^{x}$) and direct ($K^{d}$) terms
are given by (we use $i,j,\mbox{ and \ensuremath{k\dots}}$ as occupied
(hole) state indices, $a,b,\mbox{ and }c\dots$ as unoccupied (electron)
states indices, and $r,s,\mbox{ and }t\dots$ for general indices):

\begin{align}
D_{ia,bj}= & \left(\varepsilon_{a}-\varepsilon_{i}\right)\delta_{ab}\delta_{ij}\\
K_{ia,bj}^{X}= & \langle\phi_{a}\phi_{i}|\hat{v}_{C}|\phi_{b}\phi_{j}\rangle=\iint d\mathbf{r}d\mathbf{r}'\nonumber \\
\times & \phi_{i}\left(\mathbf{r}\right)\phi_{a}\left(\mathbf{r}\right)v_{C}\left(\left|{\bf r}-{\bf r}'\right|\right)\phi_{j}\left(\mathbf{r}'\right)\phi_{b}\left(\mathbf{r}'\right)\\
K_{ia,bj}^{D}= & \langle\phi_{a}\phi_{b}|\hat{W}|\phi_{i}\phi_{j}\rangle=\iint d\mathbf{r}d\mathbf{r}'\nonumber \\
\times & \phi_{b}\left(\mathbf{r}\right)\phi_{a}\left(\mathbf{r}\right)W\left(\mathbf{r},\mathbf{r}',0\right)\phi_{j}\left(\mathbf{r}'\right)\phi_{i}\left(\mathbf{r}'\right).
\end{align}
Here, $\varepsilon_{a}$ and $\varepsilon_{i}$ are the quasi-particle
energies for the virtual and occupied space (which can be obtained
from a DFT+GW calculation or from an alternative suitable approach)
and $\phi_{a}\left({\bf r}\right)$ and $\phi_{i}\left({\bf r}\right)$
are the corresponding quasi-particle orbitals; $\hat{v}_{C}$ is the
Coulomb potential while $W$ is the screened Coulomb potential, typically
calculated within the Random Phase Approximation (RPA), which can
be written in real space as:

\begin{singlespace}
\begin{equation}
W\left(\mathbf{r},\mathbf{r}',0\right)=v_{C}\left(\left|{\bf r}-{\bf r}'\right|\right)+\delta W^{RPA}\left(\mathbf{r},\mathbf{r}',0\right),\label{eq:W}
\end{equation}
with 

\begin{align}
\delta W^{RPA}\left(\mathbf{r},\mathbf{r}',0\right)= & \iint d\mathbf{r}''d\mathbf{r}'''v_{C}\left(\left|{\bf r}-\mathbf{r}''\right|\right)\nonumber \\
\tilde{\chi}^{RPA}\left({\bf \mathbf{r}}'',{\bf r}''',0\right) & \left(v_{C}\left(\left|{\bf r}'''-{\bf r}'\right|\right)+f_{XC}\mbox{\ensuremath{\left(\mathbf{r}'''\right)}}\delta\mbox{\ensuremath{\left({\bf r}'''-{\bf r}'\right)}}\right).\label{eq:deltaW}
\end{align}
Here, $f_{XC}\mbox{\ensuremath{\left(\mathbf{r}\right)}}$ is the
DFT exchange-correlation potential (if DFT is used to obtain the RPA
screening, otherwise set $f_{XC}\mbox{\ensuremath{\left(\mathbf{r}\right)}}=0$),
and $\tilde{\chi}^{RPA}\left({\bf r},{\bf r}',0\right)$ is the half-Fourier
transform (at $\omega=0$) of the real-time density-density correlation
function within the RPA (the latter can be also obtained from TDDFT,
as further discussed below). We note in passing that often the above
is solved within the Tamm-Dancoff approximation (TDA),\cite{Dyson1953,Taylor1954}
which sets $B=0$ and thus only requires the diagonalization of the
matrix $A$.
\end{singlespace}

\subsection{Time-Dependent Bethe-Salpeter Equation (TDBSE)}

The time-dependent formulation of the BSE follows from the connections
made between CIS and TDHF.\cite{Casida1995,Casida1996,Hirata1999a,Hirata1999b}
In short, solving the TDHF equations $i\hbar\frac{\partial\phi_{j}\left({\bf r},t\right)}{\partial t}=\hat{h}_{HF}\left(t\right)\phi_{j}\left({\bf r},t\right)$
for the occupied orbitals is identical to solving the symplectic eigenvalue
problem of Eq.~\eqref{eq:casida} with $\delta W\left(\mathbf{r},\mathbf{r}',0\right)=0$.
Here, $\hat{h}_{HF}=\hat{t}+\hat{v}_{ion}+\hat{v}_{H}\left(t\right)+\hat{k}_{X}\left(t\right)$
is the Hartree-Fock (HF) Hamiltonian, $\hat{t}$ is the kinetic energy,
$\hat{v}_{ion}$ is the external potential, $\hat{v}_{H}\psi\left(\mathbf{r}\right)=\int d\mathbf{r}'v_{C}\left(\left|{\bf r}-{\bf r}'\right|\right)n\left({\bf r}',t\right)\psi\left(\mathbf{r}\right)$
is the Hartree potential, and $\hat{k}_{X}\left(t\right)\psi\left(\mathbf{r}\right)=-\frac{1}{2}\int d\mathbf{r}'\rho\left(\mathbf{r},{\bf r}',t\right)v_{C}\left(\left|{\bf r}-{\bf r}'\right|\right)\psi\left(\mathbf{r'}\right)$
is the non-local exchange potential. $n\left({\bf r},t\right)=2\sum_{j}\left|\phi_{j}\left({\bf r},t\right)\right|^{2}$
and $\rho\left(\mathbf{r},{\bf r}',t\right)=2\sum_{j}\phi_{j}^{*}\left({\bf r'},t\right)\phi_{j}\left({\bf r},t\right)$
are the time-dependent electron density and density matrix, respectively.
The connection to CIS is made by realizing that for $\delta W\left(\mathbf{r},\mathbf{r}',0\right)=0$
and setting $B=0$ (the TDA), the symplectic eigenvalue problem of
Eq.~\eqref{eq:casida} is nothing else but the CIS Hamiltonian. Thus,
TDHF within the TDA and CIS are identical.

We follow a similar logic and derive an adiabatic time-dependent BSE:
\begin{equation}
i\hbar\frac{\partial\phi_{j}^{\gamma}\left({\bf r},t\right)}{\partial t}=\hat{h}_{BS}^{\gamma}\left(t\right)\phi_{j}^{\gamma}\left({\bf r},t\right)\label{eq:tdbs}
\end{equation}
where $\gamma$ is a perturbation strength (\emph{i.e.}, $\gamma=0$
is the unperturbed case, see Eq.~\eqref{eq:phi t=00003D0-1}) with
a screened effective Hamiltonian given by:
\begin{equation}
\hat{h}_{BS}^{\gamma}=\hat{h}_{qp}+\hat{v}_{H}^{\gamma}\left(t\right)-\hat{v}_{H}^{0}\left(t\right)+\hat{k}_{\epsilon X}^{\gamma}\left(t\right)-\hat{k}_{\epsilon X}^{0}\left(t\right).\label{eq:h_BS}
\end{equation}
Here, $\hat{h}_{qp}$ is the quasi-particle Hamiltonian which is typically
determined from a GW calculation correcting the quasiparticle energies
and orbitals of the underlying DFT. The GW approximation to $\hat{h}_{qp}$
is rather difficult to implement since it involves a non-local, energy-dependent
operator. An alternative is to use a DFT approach that provides an
accurate description of quasiparticle excitations.\cite{Baer2005a,Brothers2008}
However, since the exact model for $\hat{h}_{qp}$ is not the central
target of the present work, we represent it by a simple semi-empirical
local Hamiltonian of the form:\cite{Wang1994d,Wang1995,Wang1996,Fu1997,Williamson2000,Franceschetti2000a,Zunger2001}

\begin{equation}
\hat{h}_{qp}\approx\hat{t}+\hat{v}_{ps},\label{eq:h_PS}
\end{equation}
where, as before $\hat{t}$ is the kinetic energy and
$\hat{v}_{ps}=\sum_{\alpha}\hat{v}_{\alpha}$ is the empirical
pseudopotential, given as a sum of atomic pseudopotentials which were
generated to reproduce the bulk band structure, providing accurate
quasi-particle excitations in the bulk. The semiempirical approach has
been successfully applied to calculate the quasi-particle spectrum of
semiconducting nanocrystals of various sizes and
shapes.\cite{Wang1994d,Wang1996,Fu1997b,Rabani1999b,Reboredo2000,Franceschetti2000a,Franceschetti2000,Eshet2013}

In Eq.~\eqref{eq:h_BS}, $\hat{v}_{H}^{\gamma}\left(t\right)\psi\left(\mathbf{r}\right)=\int d\mathbf{r}'v_{C}\left(\left|{\bf r}-{\bf r}'\right|\right)n^{\gamma}\left({\bf r}',t\right)\psi\left(\mathbf{r}\right)$
is the Hartree potential with $n^{\gamma}\left({\bf r},t\right)=2\sum_{j}\left|\phi_{j}^{\gamma}\left({\bf r},t\right)\right|^{2}$
and $\hat{k}_{\epsilon X}^{\gamma}\left(t\right)\psi\left(\mathbf{r}\right)=-\frac{1}{2}\int d\mathbf{r}'\rho^{\gamma}\left(\mathbf{r},{\bf r}',t\right)W^{RPA}\left({\bf r},{\bf r}',0\right)\psi\left(\mathbf{r'}\right)$
is the screened exchange potential with $W^{RPA}\left({\bf r},{\bf r}',0\right)$
given by Eqs.~\eqref{eq:W} and \eqref{eq:deltaW} and $\rho^{\gamma}\left(\mathbf{r},{\bf r}',t\right)=2\sum_{j}\phi_{j}^{\gamma}\left({\bf r'},t\right)^{*}\phi_{j}^{\gamma}\left({\bf r},t\right)$.
The application of $\hat{k}_{\epsilon X}^{\gamma}\psi\left(\mathbf{r}\right)$
is further discussed below. 

\begin{figure}[H]
\centering{}\includegraphics[width=8cm]{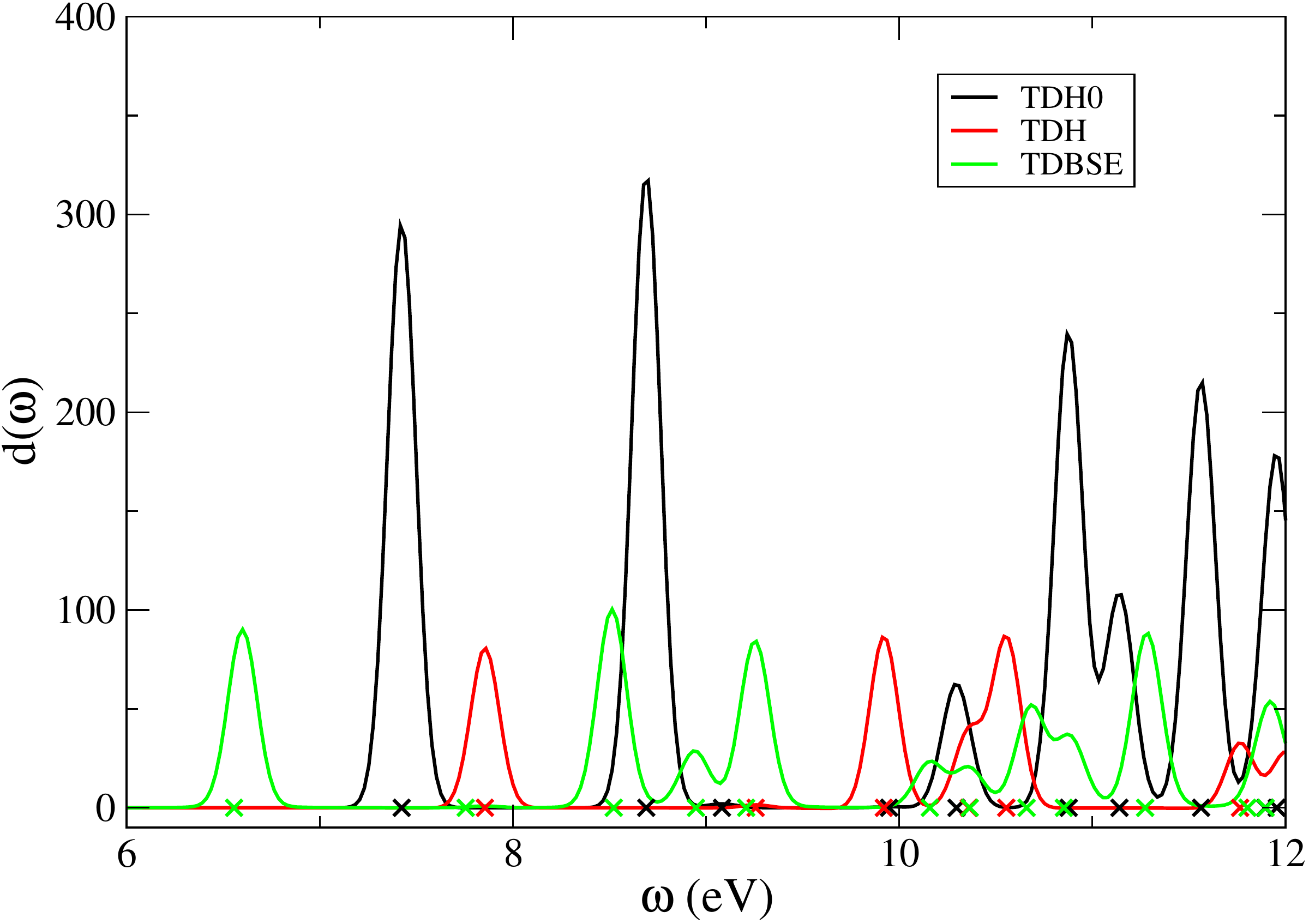}\protect\caption{\label{fig:Comparison-of-BSE}Comparison of BS calculations using
the symplectic eigenvalue (Eq.~\eqref{eq:L}, $x$-symbols) with
the frequency dependent dipole-dipole correlation generated from TDBSE
(Eqs.~\eqref{eq:h_BS}, \eqref{eq:tdbs} and \eqref{eq:h_PS}, solid
lines) for $\mbox{SiH}_{4}$. Black: TDBSE with $\gamma=0$ (TDH0)
compared with eigenvalues of Eq.~\eqref{eq:L} setting $K^{X}$ and
$K^{D}$ to zero. Red: TDBSE with $\hat{h}_{BS}^{\gamma}=\hat{h}_{qp}+\hat{v}_{H}^{\gamma}\left(t\right)-\hat{v}_{H}^{0}\left(t\right)$
(TDH) compared with Eq.~\eqref{eq:L} setting $K^{d}$ to zero (TDH).
Green: TDBSE with $\gamma=10^{-5}$au (TDBSE) compared with Eq.~\eqref{eq:L},
both for $\epsilon=5$. }
\end{figure}

In analogy with the relations derived between TDHF and its eigenvalue
representation, it is clear that the time-dependent formulation for
the BSE given by Eqs.~\eqref{eq:tdbs} and \eqref{eq:h_BS} is identical
to the full symplectic eigenvalue problem of Eq.~\eqref{eq:casida}.
In Fig.~\ref{fig:Comparison-of-BSE} we compare the results for
$\mbox{SiH}_{4}$ on a $8\times8\times8$ grid generated by propagating
the occupied orbitals with the Bethe-Salpeter Hamiltonian
\eqref{eq:h_BS} (TDBSE) to the exact diagonalization of
Eq.~\eqref{eq:L} (static approach).  We use a local semi-empirical
pseudopotential that has been applied successfully to study the
optical properties of silicon
nanocrystals.\cite{Wang1994d,Wang1994c,Zunger1996} For both the direct
approach and the TDBSE we approximate $W\left({\bf r},{\bf
  r}',0\right)$ by $\epsilon^{-1}v_{C}\left(\left|{\bf r}-{\bf
  r}'\right|\right)$, where $\epsilon$ is a constant screening
parameter. The idea is to confirm that the eigenvalues of
Eq.~\eqref{eq:L} and the time-dependent version of the BSE are
identical (validating both the theory and the implementation).

The time-domain calculations are based on a linear-response approach
to generate the dipole-dipole correlation function $d\left(t\right)$
and its Fourier transform $\tilde{d}\left(\omega\right)=\int_{0}^{\infty}dt\, e^{i\omega t}d\left(t\right)$.
In short, we perturb the occupied eigenstates ($\phi_{j}\left(\mathbf{r}\right)$)
of $\hat{h}_{qp}$ at $t=0$:

\begin{equation}
\phi_{j}^{\gamma}\left({\bf r},t=0\right)=e^{-i\gamma z/\hbar}\phi_{j}\left({\bf r}\right),\label{eq:phi t=00003D0-1}
\end{equation}
where for simplicity, we assume that the dipole is in the $z$--direction.
We then propagate these orbitals according to Eq.~\eqref{eq:tdbs}
and generate the dipole-dipole correlation function:

\begin{equation}
d\left(t\right)=\frac{1}{\gamma}\int d\mathbf{r}\, z\left(n^{\gamma}\left({\bf r},t\right)-n^{0}\left({\bf r},t\right)\right),\label{eq:d-1}
\end{equation}
where as before $n^{\gamma}\left({\bf r},t\right)=2\sum_{j}\left|\phi_{j}^{\gamma}\left({\bf r},t\right)\right|^{2}$and
$\gamma$ is a small parameter representing the strength of the perturbation,
typically $10^{-3}-10^{-5}\hbar E_{h}^{-1}$. 

The agreement for the position of the excitations (solid lines) generated
by the time-domain BSE is perfect with the static calculation ($x$-symbols),
as seen in Fig.~\ref{fig:Comparison-of-BSE}. The resolved individual
transitions are broadened reflecting the finite propagation time used
for the time-domain calculations. We find that in some cases the oscillator
strength is very small and thus a transition is not observed in $\tilde{d}\left(\omega\right)$.

\begin{figure}[H]
\centering{}\includegraphics[width=8cm]{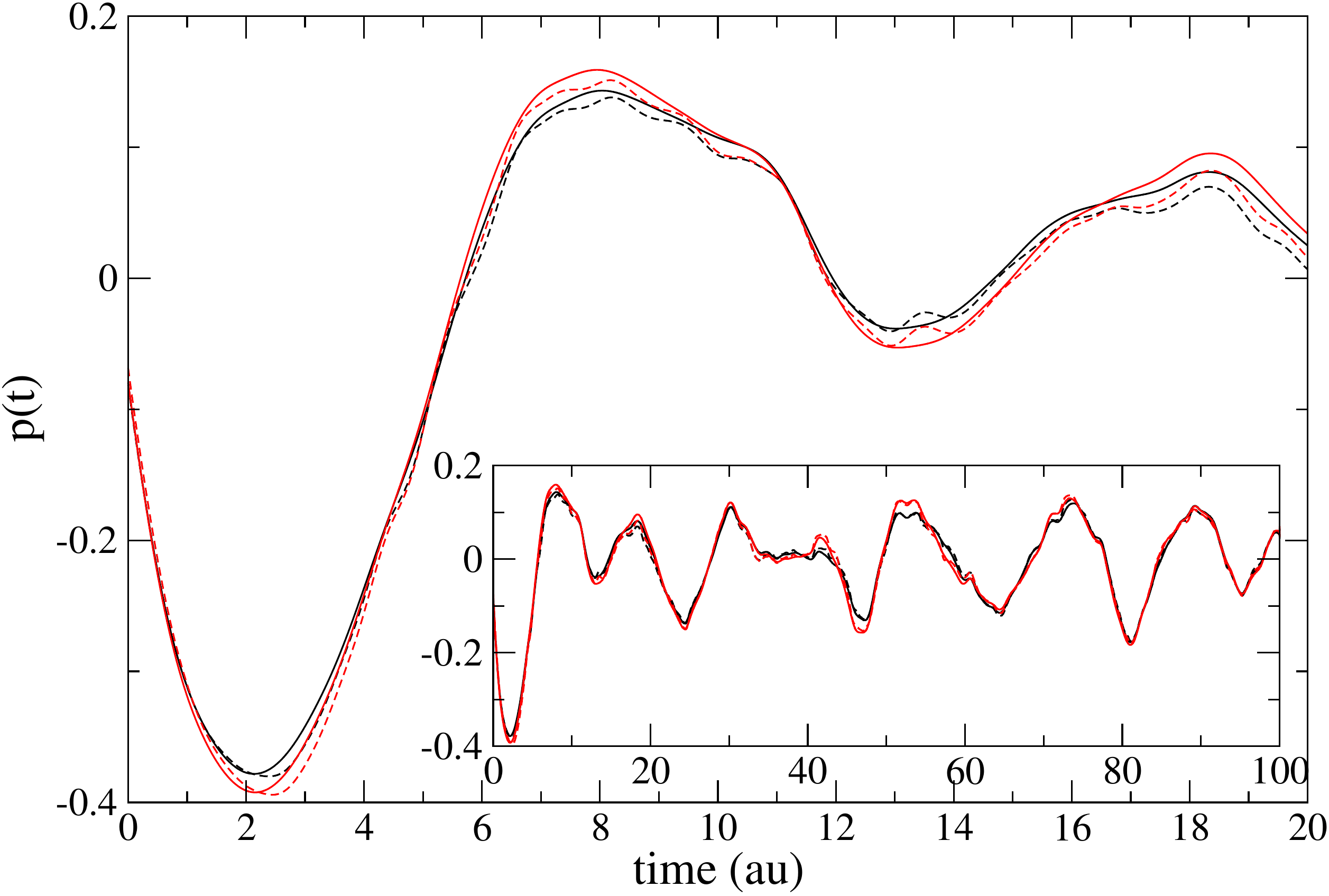}\protect\caption{\label{fig:Average-momentum-calculated}Average momentum along the
$z$-direction calculated in two ways (see text for more details)
for $\mbox{SiH}_{4}$ using the TDH (black curves) and the TDBSE (red
curves) methods. Solid and dashed curves where generated using the
expectation value of the momentum (Eq.~\eqref{eq:p1}) and the numerical
derivative of the expectation value of the position (Eq.~\eqref{eq:p2}),
respectively. Inset: same for longer times.}
\end{figure}

An additional important test of the TDBSE formalism is whether the
Hamiltonian in Eq.~\eqref{eq:h_BS} preserves the Ehrenfest theorem
(see Appendix B for more details). Naturally, this would be the case
if $\hat{h}_{qp}$ would include the terms $\hat{v}_{H}^{0}\left(t\right)$
and $\hat{k}_{\epsilon X}^{0}\left(t\right)$, such that they cancel
out for $\hat{h}_{BS}^{\gamma}.$ However, for an arbitrary choice
of $\hat{h}_{qp}$ this needs to be confirmed. In Figure~\ref{fig:Average-momentum-calculated}
we plot the average momentum for $\mbox{SiH}_{4}$ calculated in two
different ways. The solid curves were obtained directly from:
\begin{equation}
\frac{\left\langle \mathbf{p}\left(t\right)\right\rangle }{m}=-2i\hbar\sum_{j}\int d\mathbf{r}\phi_{j}^{\gamma}\left(\mathbf{r},t\right)^{*}\frac{\partial}{\partial\mathbf{r}}\phi_{j}^{\gamma}\left(\mathbf{r},t\right),\label{eq:p1}
\end{equation}
while the dashed curves were obtained by taking the numerical time
derivative (central difference) of the expectation value of $\mathbf{r}\left(t\right):$
\begin{equation}
\frac{\left\langle \mathbf{p}\left(t\right)\right\rangle }{m}=\frac{\partial}{\partial t}\left\langle \mathbf{q}\left(t\right)\right\rangle =2\frac{\partial}{\partial t}\sum_{j}\int d\mathbf{r}\phi_{j}^{\gamma}\left(\mathbf{r},t\right)^{*}\mathbf{q}\phi_{j}^{\gamma}\left(\mathbf{r},t\right).\label{eq:p2}
\end{equation}
The agreement is not perfect but improves with decreasing the time
step $\delta t$ (not shown here). We also show the results for the
time-dependent Hartree (TDH), i.e., ignoring the screened exchange
term in $\hat{h}_{BS}^{\gamma}$. The deviations observed for TDBSE
and TDH are similar, although for TDH the Ehrenfest theorem holds
exactly and thus the agreement should be perfect. The difference are
associated with numerical inaccuracies resulting from the finite time
step and grid used in the calculation. The inset shows that the deviations
are insignificant even at much longer times over many periods.

\section{Time-Dependent Stochastic Bethe-Salpeter Equation\label{sec:tdsbse} }

We consider two formulations for the time-dependent stochastic BSE
(TD\emph{s}BSE). The first approach is a direct generalization of the
approach we have recently developed for the stochastic
TDH,\cite{Gao2014TDsDFT} in which we describe an efficient way to
account for the the screened exchange term in the
$\hat{h}_{BS}^{\gamma}$. This approach works well for short times,
however, unlike in TDH, the inclusion of an exchange term requires an
increasing number of stochastic orbitals with the system size. The
second approach offers access to timescales relevant for most
spectroscopic applications at a practical quadratic computational
cost.

\subsection{Extending the Stochastic TDH to Include a Screened Exchange Term}

\begin{figure*}[t]
\centering{}\includegraphics[width=12cm]{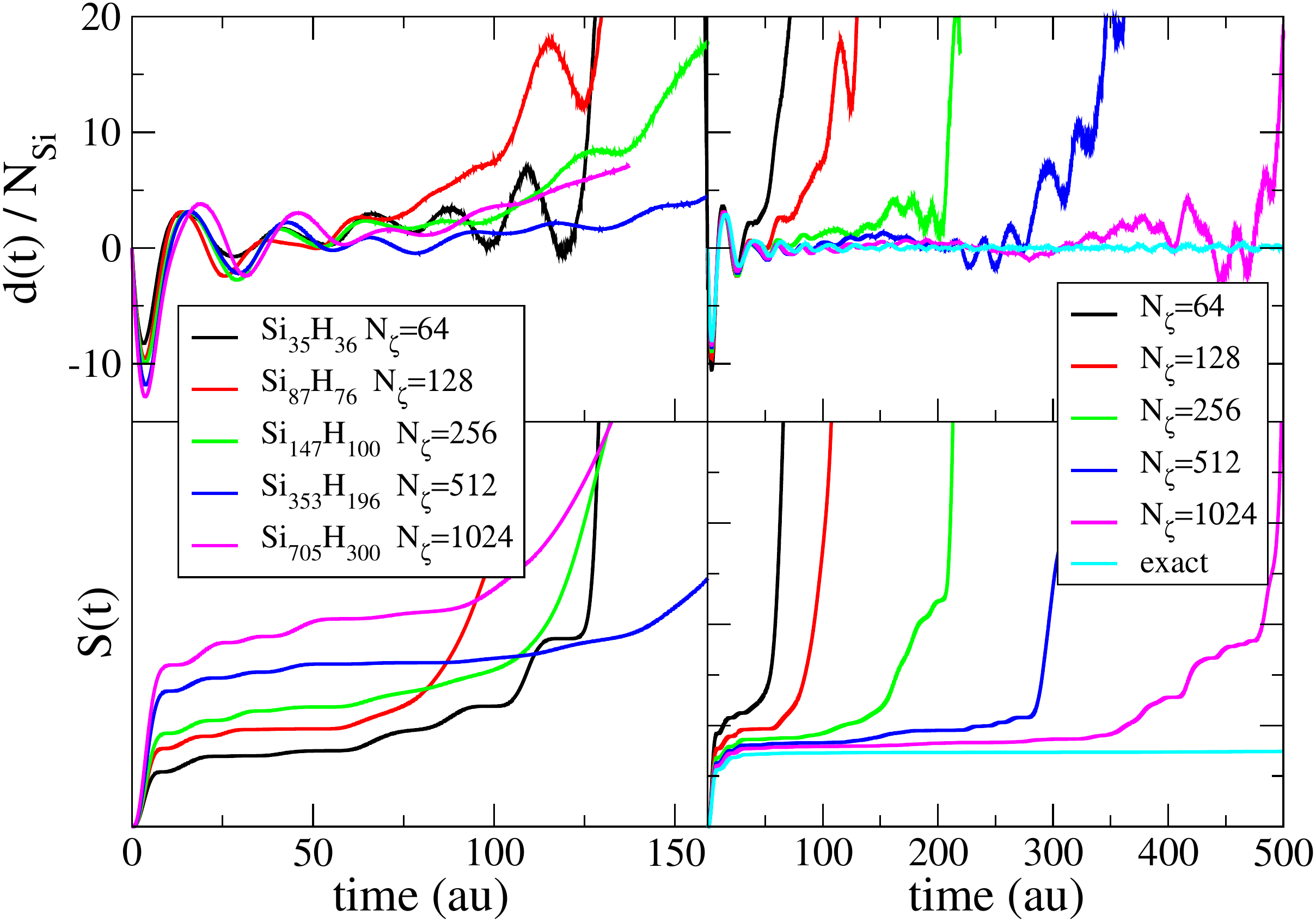}\protect\caption{\label{fig:convergence} Upper left: Dipole-dipole correlation function
(Eq.~\eqref{eq:d-1}) normalized to the number of silicon atoms in
the nanocrystal ($N_{Si}$) for several nanocrystals sizes. For each
size we use a different number of stochastic orbitals. Lower left:
Integrate dipole-dipole correlation $S\left(t\right)=\int_{0}^{t}ds\, d\left(s\right)^{2}$.
The onset of divergence scales roughly linearly with the size. Upper
right: Dipole-dipole correlation function normalized to the number
of silicon atoms for $\mbox{Si}_{87}\mbox{H}_{76}$ for different
values of $N_{\zeta}$. Lower right: Corresponding values for $S\left(t\right)$.}
\end{figure*}

We limit the discussion, in the body of this paper, to the case where
$W\left({\bf r},{\bf r}',0\right)$ is replaced by $\epsilon^{-1}v_{C}\left(\left|{\bf r}-{\bf r}'\right|\right)$,
where $\epsilon$ is a function of $\left|{\bf r}-{\bf r}'\right|$.
The algorithm for the TD\emph{s}BSE is based on the following steps:
\begin{enumerate}
\item Generate $N_{\zeta}$ stochastic orbitals $\zeta_{j}\left({\bf r}\right)=e^{i\theta_{j}\left({\bf r}\right)}/\sqrt{\delta V}$,
where $\theta_{j}\left({\bf r}\right)$ is a uniform random variable
in the range $\left[0,2\pi\right]$ at each grid point (total of $N_{g}$
grid points), $\delta V$ is the volume element of the grid, and $j=1\,,\dots,\, N_{\zeta}$.
The stochastic orbitals obey the relation $\mathbf{1}=\left<\left|\zeta\left\rangle \right\langle \zeta\right|\right>_{\zeta}$
where $\left<\cdots\right>_{\zeta}$ denotes a statistical average
over $\zeta$.
\item Project each stochastic orbital $\zeta_{j}\left({\bf r}\right)$ onto
the occupied space: $\left|\xi_{j}\right\rangle \equiv\sqrt{\theta_{\beta}\left(\mu-\hat{h}_{qp}\right)}\left|\zeta_{j}\right\rangle $,
where $\theta_{\beta}\left(x\right)=\frac{1}{2}\mbox{erfc}\left(\beta\left(x\right)\right)$
is a smooth representation of the Heaviside step function~\cite{Baer2013}
and $\mu$ is the chemical potential. The action of $\sqrt{\hat{\theta}_{\beta}}$
is performed using a suitable expansion in terms of Chebyshev polynomials
in the static quasi-particle Hamiltonian with coefficients that depend
on $\mu$ and $\beta$.\cite{Kosloff1988} 
\item Define non-perturbed and perturbed orbitals for $t=0$ to the orbitals:
$\xi_{j}^{0}\left({\bf r},t=0\right)=\xi_{j}\left({\bf r}\right)$,
$\xi_{j}^{\gamma}\left({\bf r},t=0\right)=e^{-iv\left({\bf r}\right)/\hbar}\xi_{j}\left({\bf r}\right)$.
For the absorption spectrum, the perturbation is given by $v\left(\mathbf{r}\right)=r_{\alpha}$
and $\alpha\equiv x,y,z$. 
\item Propagate the perturbed ($\xi_{j}^{\gamma}\left({\bf r},t\right)$)
and unperturbed ($\xi_{j}^{0}\left({\bf r},t\right)$) orbitals according
to the adiabatic time-dependent BSE: 
\begin{align}
i\hbar\frac{\partial\xi_{j}^{\gamma}\left({\bf r},t\right)}{\partial t} & =\hat{h}_{BS}^{\gamma}\left(t\right)\xi_{j}^{\gamma}\left({\bf r},t\right).\label{eq:tdsdft-2}
\end{align}
Use the split operator technique to perform the time propagation from
time $t$ to time $t+\Delta t$: 
\begin{align}
e^{-\frac{i}{\hbar}\hat{h}_{BS}^{\gamma}\Delta t} & \approx e^{-\frac{i}{2\hbar}\left(\hat{v}_{ps}+\hat{v}_{H}^{\gamma}\left(t\right)-\hat{v}_{H}^{0}\left(t\right)\right)\Delta t}\nonumber \\
\times & e^{-\frac{i}{2\hbar}\hat{t}\Delta t}e^{-\frac{i}{\hbar}\left(\hat{k}_{\epsilon X}^{\gamma}\left(t\right)-\hat{k}_{\epsilon X}^{0}\left(t\right)\right)\Delta t}\nonumber \\
\times & e^{-\frac{i}{2\hbar}\hat{t}\Delta t}e^{-\frac{i}{2\hbar}\left(\hat{v}_{ps}+\hat{v}_{H}^{\gamma}\left(t\right)-\hat{v}_{H}^{0}\left(t\right)\right)\Delta t}
\end{align}
where propagator step involving the non-local screened exchange is
applied using a Taylor series (in all applications below we stop at
):
\begin{align}
e^{-\frac{i}{\hbar}\left(\hat{k}_{\epsilon X}^{\gamma}\left(t\right)-\hat{k}_{\epsilon X}^{0}\left(t\right)\right)\Delta t} & \approx\nonumber \\
1- & \frac{i}{\hbar}\left(\hat{k}_{\epsilon X}^{\gamma}\left(t\right)-\hat{k}_{\epsilon X}^{0}\left(t\right)\right)\Delta t+\cdots
\end{align}

\item The application of $\hat{h}_{BS}^{\gamma}\left(t\right)$ is done
as follows:

\begin{enumerate}
\item The kinetic energy is applied using a Fast Fourier Transform (FFT). 
\item The Hartree term is generated using convolution and FFT with the density
obtained from the stochastic orbitals: 
\begin{align}
n^{\gamma}\left({\bf {\bf r}},t\right) & =\frac{2}{N_{\zeta}}\sum_{j=1}^{N_{\zeta}}\left|\xi_{j}^{\gamma}\left({\bf r},t\right)\right|^{2}.\label{eq:dens-from-xi-3}
\end{align}

\item The time-consuming part of the application of $\hat{h}_{BS}^{\gamma}$
on a vector $\psi$ in Hilbert space is $\hat{k}_{\epsilon X}^{\gamma}\left(t\right)-\hat{k}_{\epsilon X}^{0}\left(t\right)$.
This operation scales as $O\left(NN_{grid}\right)$ and one needs
to carry this for all occupied states, leading a $O\left(N^{2}N_{grid}\right)$
computational scaling. To reduce this high scaling resulting from
the exchange operation we use the same philosophy underlying this
work, \emph{i.e.}, replacing summation with stochastic averaging.
In practice we therefore replace the summation over occupied orbitals
in the exchange operation by acting with very few $n_{\eta}\ll N_{\zeta}$,
typically $n_{\eta}=1-16,$ stochastic orbitals write the exchange
operation as:
\begin{align}
\hat{k}_{\epsilon X}^{\gamma}\left(t\right)\psi\left(\mathbf{r},t\right) & =\frac{1}{n_{\eta}}\sum_{x=1}^{n_{\eta}}\eta_{x}^{\gamma}\left(\mathbf{r},t\right)\nonumber \\
\times\mbox{\ensuremath{\int}d\textbf{r}}'\epsilon^{-1} & v_{C}\left(\left|{\bf r}-{\bf r}'\right|\right)\eta_{x}^{\gamma}\left(\mathbf{r}',t\right)^{*}\psi\left(\mathbf{r}',t\right).\label{eq:stoch exchange-1}
\end{align}
The key is that these stochastic orbitals are defined as a different
random combination of the full set of orbitals at any given time step$\eta_{x}^{\gamma}$
are defined as random superpositions of the $N_{\zeta}$ stochastic
orbitals:
\begin{align}
\eta_{x}^{\gamma}\left(\mathbf{r},t\right) & =\frac{1}{N_{\zeta}}\sum_{j}^{N_{\zeta}}e^{i\alpha_{xj}(t)}\xi_{j}^{\gamma}\left({\bf r},t\right).
\end{align}
To improve the representation of the stochastic exchange operators,
the random phases $\alpha_{xj}\left(t\right)$ are re-sampled at each
time step. Note that the same phases are used for both $\eta_{x}^{\gamma}\left(\mathbf{r},t\right)$
and $\eta_{x}^{0}\left(\mathbf{r},t\right)$. This use of stochastic
orbitals reduces the overall scaling of the method to quadratic, since
$n_{\eta}$ does not dependent on the system size.
\end{enumerate}
\end{enumerate}
In Fig.~\ref{fig:convergence} we show the calculated $d\left(t\right)$
and $S\left(t\right)=\int_{0}^{t}ds\, d\left(s\right)^{2}$ for a
series of silicon nanocrystals. We used $n_{\zeta}=16$ which leads
to results that are indistinguishable from $n_{\zeta}=N_{\zeta}$
(though even a smaller $n_{\zeta}$would have been sufficient). We
used a constant value for$\epsilon=5$ and the time step was $\Delta t=0.025\mbox{au}$.

In general, we find that the results converge up to a time $\tau_{C}$
and then the signal diverges exponentially. Several conclusions can
be drawn from these calculations:
\begin{enumerate}
\item The stochastic approximation to $d\left(t\right)$ oscillates about
zero up to a time $\tau_{C}$, but this is followed by a gradual increase
which eventually leads to divergence (upper panels of Fig.~\ref{fig:convergence}).
\item $\tau_{C}$ increases with the number of stochastic orbitals, $N_{\zeta}$,
roughly as $\tau_{C}\propto N_{\zeta}^{\alpha}$ with $\alpha=1-2$
(right panels of Fig.~\ref{fig:convergence}). This is somewhat better
than the case for TDH for which $\tau_{C}$ roughly scaled as $N_{\zeta}^{1/2}$.
\item $\tau_{C}$ decreases with increasing system size roughly as $\frac{1}{N_{e}}$,
where $N_{e}$ is the number of electrons (left panels of Fig.~\ref{fig:convergence}).
Therefore, to converge the results to a fixed $\tau_{C}$ one has
to increase $N_{\zeta}$ roughly linearly with the system size . This
leads to a quadratic scaling of the approach. In TDH the opposite
is true, $\tau_{C}$ \emph{increases} with increasing system size
due to self-averaging.\cite{Gao2014TDsDFT}
\item To reach times sufficient for most spectroscopic applications, the
number of stochastic orbitals exceeds that of occupied states ($N_{\zeta}>N_{occ})$.
\end{enumerate}
To conclude this subsection, we find that this version of a TD\emph{s}BSE
scales roughly quadratically with the system size, rather than sub-linearly
for TDH. Furthermore, to calculate the response to meaningful times,
the naive extention of the TDH to include exchange requires a rather
large number of stochastic orbitals ($N_{\zeta}$), often much larger
than the number of occupied orbitals. However, it is sufficient to
represent the operation of the exchange Hamiltonian with a relatively
small set of linear combination of all stochastic orbitals ($n_{\zeta}$).
We next show how the method can be improved significantly increasing
$\tau_{C}$ to values much larger than required to obtain the spectrum
in large systems.

\subsection{Time-Dependent \textit{Stochastic} Bethe-Salpeter with Orthogonalization}

To circumvent the pathological behavior observed above, we propose
to orthogonalize the projected stochastic orbitals (after step ``2'').
This requires that $N_{\zeta}$ be equal to the number of occupied
states $N_{occ}$. However, this makes the TD\emph{s}BSE stable for
time-scale exceeding $50$fs, which for any practical spectroscopic
application for large systems is more than sufficient. Formally, since
the number of stochastic orbitals (equal to the number of occupied
states) increases linearly with the system size, the approach scales
as $O\left(N_{\zeta}N_{g}\right)$. The orthogonalization step scales
formally as $O\left(N_{\zeta}^{2}N_{g}\right)$, however, for the
size of systems studied here, it is computationally negligible compared
with the projection and propagation steps.

\begin{figure*}[t]
\centering{}\includegraphics[width=16cm]{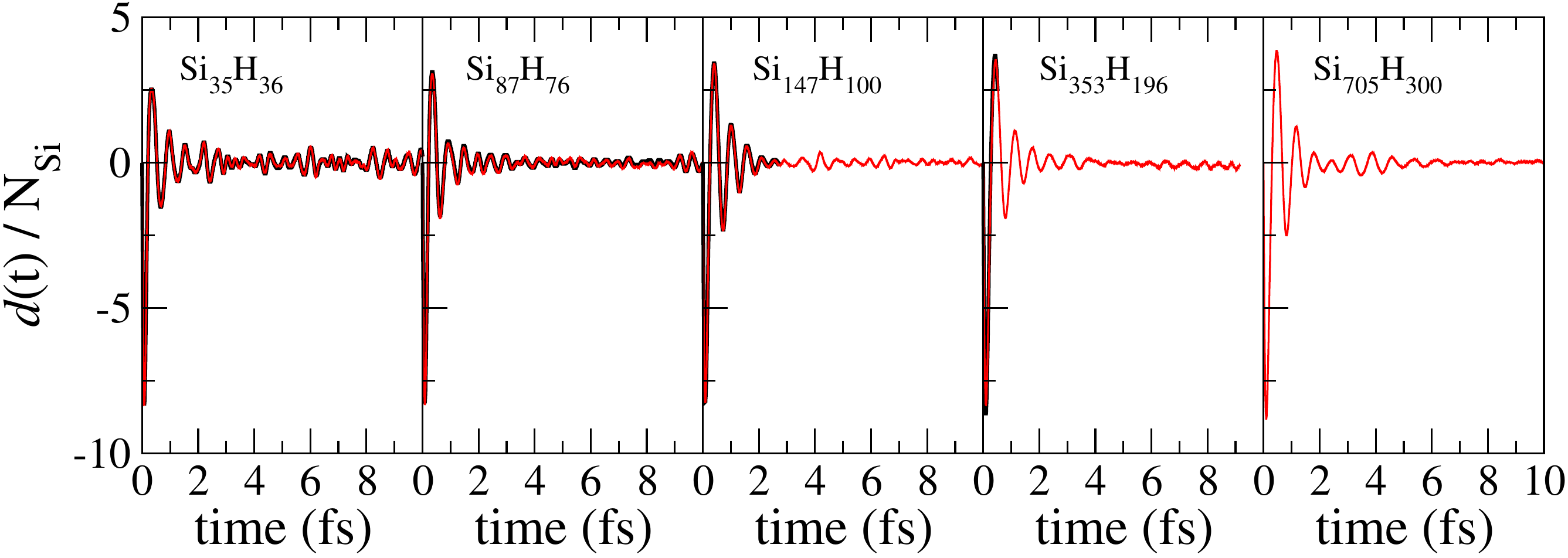}\protect\caption{\label{fig:dip} The dipole-dipole correlation function calculated
using the TD\textit{s}BSE approach with orthogonalization (red curves)
compared with a direct time-dependent BSE approach (black curves).
Note that the direct (i.e., non-stochastic) BSE approach is so expensive
due to the full-exchange operation that it was not done for the largest
NCs and was only followed for short times for intermediate size NCs.}
\end{figure*}

In Fig.~\ref{fig:dip} we compare the dipole-dipole correlation function
computed from the TD\textit{s}BSE with $n_{\zeta}=1$ to the direct
TDBSE approach for silicon nanocrystals of varying sizes ($\mbox{Si\ensuremath{_{35}\mbox{H}_{36}}}$,
$\mbox{Si\ensuremath{_{87}\mbox{H}_{76}}}$, $\mbox{Si\ensuremath{_{147}\mbox{H}_{100}}}$,
$\mbox{Si\ensuremath{_{353}\mbox{H}_{196}}}$, and $\mbox{Si\ensuremath{_{705}\mbox{H}_{300}}}$).
The purpose is to demonstrates the power of the TD\textit{s}BSE approach
with orthogonalization. Therefore, for simplicity $W\left({\bf r},{\bf r}',0\right)$
is replaced by $\epsilon^{-1}v_{C}\left(\left|{\bf r}-{\bf r}'\right|\right)$
with $\epsilon=5$ for all system sizes. Clearly, even when $n_{\zeta}=1$,
the TD\textit{s}BSE is in perfect agreement with the direct TDBSE
approach. The cubic scaling of the later limits the application to
small NCs or to short times. 

\begin{figure*}[t]
\begin{centering}
\includegraphics[width=16cm]{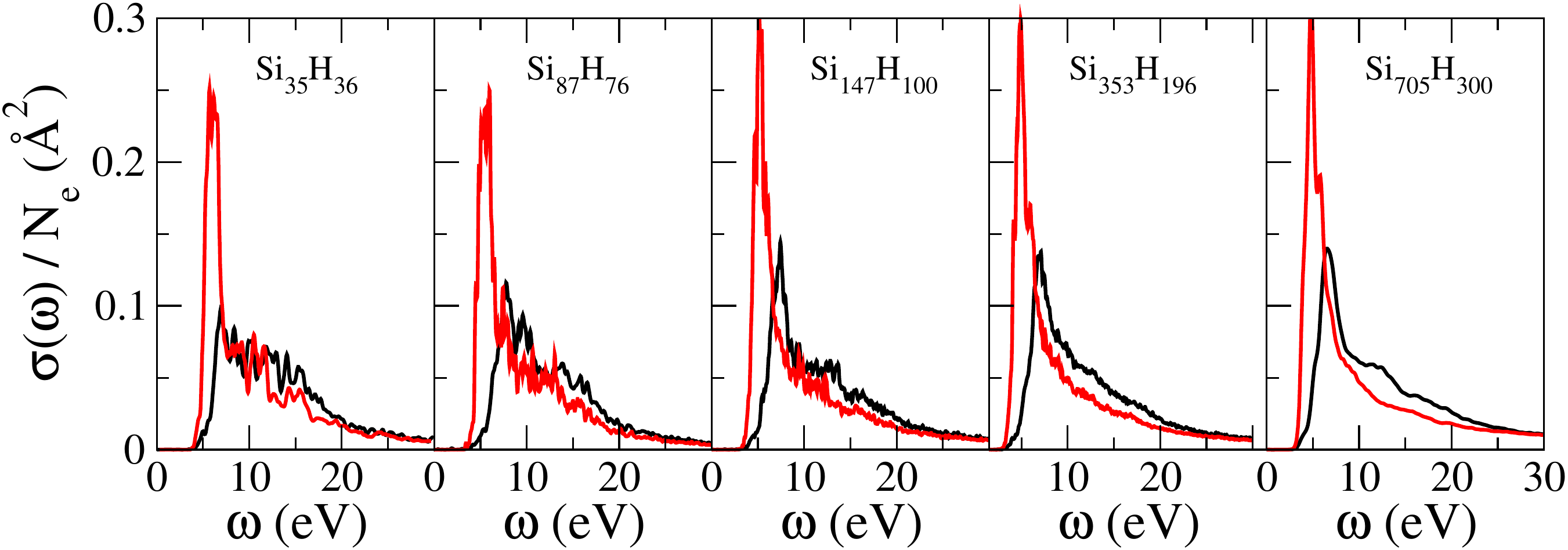}
\par\end{centering}

\centering{}\includegraphics[width=16cm]{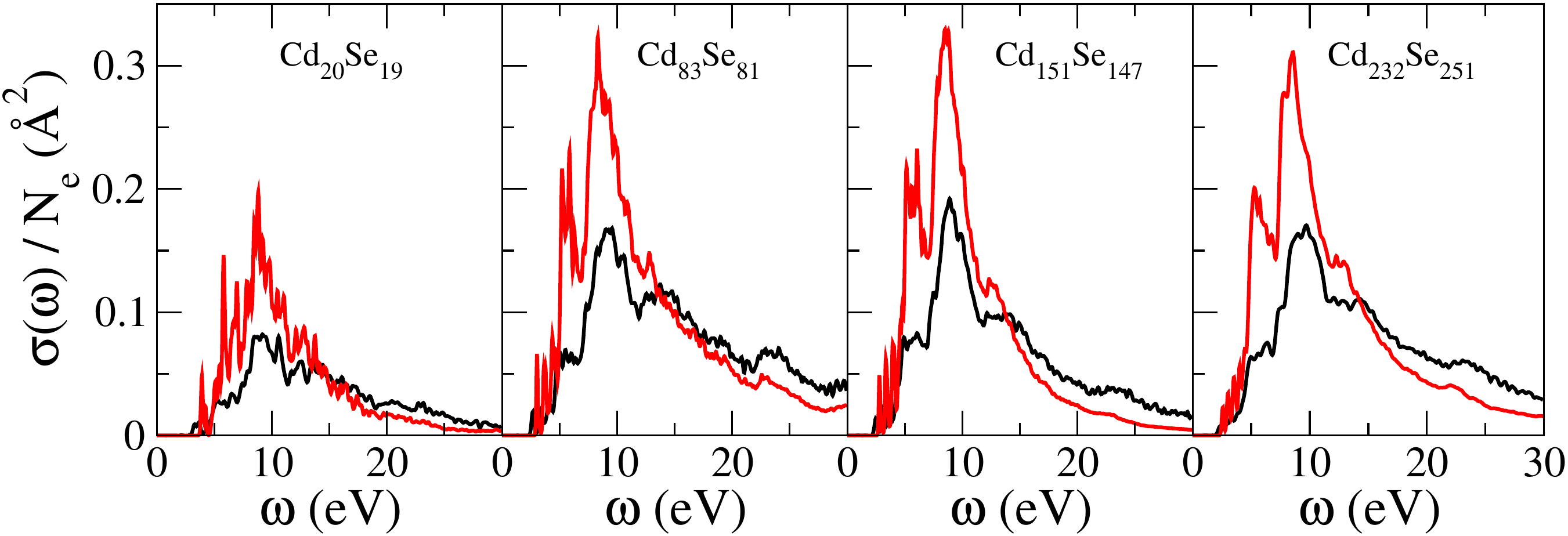}\protect\caption{\label{fig:spec} Upper panels: The absorption cross section for silicon
NCs computed by Fourier transforming the TD\textit{s}BSE dipole-dipole
correlation function (black curves) and the corresponding absorption
cross section computed for $\hat{h}_{qp}$, \emph{i.e., }by ignoring
electron-hole interactions.}
\end{figure*}

In Fig.~\ref{fig:spec} we plot the TD\textit{s}BSE absorption cross
section
($\sigma\left(\omega\right)=\frac{e^{2}}{\epsilon_{0}c}\omega\int
d{\bf r}d{\bf r}'\, z\,{\bf \tilde{\chi}\left({\bf r},{\bf
    r}',\omega\right)}\, z'$) compared to the absorption cross section
computed by ignoring the electron-hole interactions for a wide range
of energies. It is practically impossible to obtain the absorption
cross section over this wide energy range by a direct diagonalization
of the symplectic eigenvalue equation (cf.,
Eq.~\eqref{eq:casida}). Thus, so far the BSE has been applied to
relatively small nanocrystals or by converging only the low lying
excitonic transitions, even within the crude approximation where
$W\left({\bf r},{\bf r}',0\right)$ is replaced by
$\epsilon^{-1}v_{C}\left(\left|{\bf r}-{\bf r}'\right|\right)$.  As
far as we know the results shown in Fig.~\ref{fig:spec} are the first
to report a converged BS calculations for NCs of experimentally
relevant sizes. We used a constant $\epsilon$ in each run, with values
of $5$, $6.2$, $7$, $8.2$, and $8.8$ taken from
Ref.~\onlinecite{Wang1994c} for the silicon NCs (in ascending order)
and $4.5$, $5$, $5.2$ and $5.4$ for the CdSe NCs taken from
Ref.~\onlinecite{Wang1996}. The inclusion of a more accurate
description of the screening as proposed in detailed in Appendix A is
left open for future study.

For both types of NCs there is a shift of the onset of absorption to
lower energies with increasing NC size due to the quantum confinement
effect. The absorption cross section of the smallest NCs is
characterized by detailed features, which are broadened and eventually
washed out as the NC size increases. For silicon NCs, the
semi-empirical pseudopotential model over-emphasizes the lowest
excitonic transition in comparison to the plasmonic resonance observed
at $\sim10{\rm eV}$ using
TDDFT.\cite{Chelikowsky2003,Tiago2006,Ramos2008,Gao2014TDsDFT} It also
misses the split of the lowest excitonic peak observed experimentally
for bulk silicon and reproduced by the BSE
approach,\cite{Rohlfing2000,Sottile2007,Ramos2008,Rocca2012a} but not
by the current model ignoring electron-hole
correlations.\cite{Wang1994c} The fact that the current calculation
does not capture this split could be a consequence of the
approximation used to model the screening.

The results for silicon NCs seem to imply that the inclusion of electron-hole
interactions leads to a blue shift in the absorption cross section
(black curve is shifted to higher energies compared to the red curve).
Since silicon is an indirect band gap material, the onset of absorption
is not a good measure of the strength of the electron-hole interactions.
Indeed, when the approach is applied to CdSe NCs (lower panels of
Fig.~\ref{fig:spec}) the inclusion of electron-hole interaction
clearly shifts the onset of absorption to lower energies.

\section{Conclusions\label{sec:conclusions} }

We have developed a real-time stochastic approach to describe electron-hole
excitations in extended finite systems based on the BSE. Following
the logic connecting TDHF and CIS, we showed that a solution to a
Schrödinger-like time-dependent equation for the quasiparticle orbitals
with an effective Hamiltonian containing both direct and screened
exchange terms is equivalent to the symplectic eigenvalue representation
of the BSE. A direct solution of the TDBSE leads to at least cubic
scaling with the system size due to the need to compute all occupied
quasiparticle orbitals and the complexity of applying the screened
exchange term to preform the time propagation. The lower bound is
similar to the scaling of the TDHF method and thus, limits the application
of the TDBSE approach to relatively small systems. To overcome this
bottleneck, we developed a stochastic approach inspired by our previous
work on stochastic GW~\cite{Neuhauser2014a} (\textit{s}GW) and stochastic
TDDFT,\cite{Gao2014TDsDFT} in which the occupied quasiparticle
orbitals were replaced with stochastic orbitals. The latter were then
used to obtain both the RPA screening using the approach developed
for the screening in \textit{s}GW and the exchange potential by extending
the approach used todescribe the Hartree term in TD\textit{s}DFT.
Both the RPA screening the application of the exchange potential scale
nearly linearly with system size (as opposed to quadratic scaling
for example for the exchange potential). The number of stochastic
orbitals required to converge the calculation scales with system size
and thus, the overall scaling of the TD\textit{s}BSE approach is quadratic
(excluding the cubic contribution from the orthogonalization of the
stochastic orbitals, which for the system sizes studied here is a
negligible step).

We have applied the TD\textit{s}BSE approach to study optical excitations
in a wide range of energies (up to $30$~eV) in silicon and CdSe
nanocrystals with sizes up to $\approx3000$ electrons ($\approx3$~nm
diameter) and compared the results with the quasiparticle excitation
spectrum obtained within the semi-empirical pseudopotential approach.
For both systems, we find that including electron-hole correlations
broadens the spectral features and shifts the oscillator strength
to higher energies due to amplification of a plasmon resonance near
$10$~eV. For silicon we find a surprising result where the onset
optical excitations seem to shift to higher energies compared to the
quasiparticle excitations. This is a result of two factors. First,
silicon is an indirect band gap material and th the onset of optically
allowed transitions is above the the lowest excitonic state. Second,
the inclusion of electron-hole interactions via the BSE leads to an
amplification of a plasmon resonance at $\approx10$~eV shifting
the oscillator strength to higher energies at the expense of the lower
frequency absorption. These combined effects lead to an apparent shift
of the absorption onset to higher energies when electron-hole interactions
are included. This is not the case for CdSe, where the onset of optical
excitation is below the onset of the quasiparticle excitation, as
expect for a direct band-gap material.

The TD\textit{s}BSE provides a platform to obtain optical excitations
in extended systems covering a wide energy range. To overcome the
divergent behavior at long times, it is necessary to increase the
number of stochastic orbitals as the size of the system increases.
We are working in improvements of this flaw and if solved, an even
faster, linear scaling BS approach will emerge. This and other improvements
as well as more general applications will be presented in a future
work.
\begin{acknowledgments}
RB and ER are supported by The Israel Science Foundation -- FIRST
Program (grant No. 1700/14). D. N. acknowledges support by the National
Science Foundation (NSF), Grant CHE-1112500. 
\end{acknowledgments}

\subsection*{Appendix A: RPA screened exchange for TD\textit{s}BSE}

The above approach assumes that $W\left({\bf r},{\bf r}',0\right)=v_{C}\left(\left|{\bf r}-{\bf r}'\right|\right)+\delta W_{RPA}\left(\mathbf{r},\mathbf{r}',0\right)$
is approximated by $\epsilon^{-1}v_{C}\left(\left|{\bf r}-{\bf r}'\right|\right)$.
In typical BS applications, one uses the RPA screening to describe
$W\left({\bf r},{\bf r}',0\right)=v_{C}\left(\left|\mathbf{r}-\mathbf{r}'\right|\right)+\delta W\left(\mathbf{r},\mathbf{r}',0\right)$.
The stochastic formalism, however, furnishes a potentially viable
approach to overcome the assumption made to obtain $W\left({\bf r},{\bf r}',0\right)$
in this work. In the linear response limit, $\delta W_{RPA}\left(\mathbf{r},\mathbf{r}',0\right)$
can be written as: 
\begin{align}
\delta W_{RPA}\left(\mathbf{r},\mathbf{r}',0\right) & =\iint d\mathbf{r}''d\mathbf{r}'''v_{C}\left(\left|{\bf r}-\mathbf{r}''\right|\right)\nonumber \\
 & \tilde{\chi}_{RPA}\left({\bf \mathbf{r}}'',{\bf r}''',0\right)\times\nonumber \\
 & \left(v_{C}\left(\left|{\bf r}'''-{\bf r}'\right|\right)+f_{XC}\mbox{\ensuremath{\left(\mathbf{r}'''\right)}}\delta\mbox{\ensuremath{\left({\bf r}'''-{\bf r}'\right)}}\right),\label{eq:deltaW_2}
\end{align}
and we are concerned with the application of $\hat{k}_{\epsilon X}^{\gamma}\left(t\right)$
on $\psi\left(\mathbf{r},t\right)$, or more accurately, the portion
that depends on the screening:
\begin{equation}
\delta\hat{k}_{\epsilon X}^{\gamma}\left(t\right)=\eta_{x}^{\gamma}\left(\mathbf{r},t\right)\mbox{\ensuremath{\int}d\textbf{r}}'\delta W_{RPA}\left({\bf r},{\bf r}',0\right)\eta_{x}^{\gamma}\left(\mathbf{r}',t\right)^{*}\psi\left(\mathbf{r}',t\right).\label{eq:delta W_3}
\end{equation}
We first insert Eq.~\eqref{eq:deltaW} into Eq.~\eqref{eq:delta W_3}:
\begin{align}
\delta\hat{k}_{\epsilon X}^{\gamma}\left(t\right)\psi\left(\mathbf{r},t\right) & =\eta_{x}^{\gamma}\left(\mathbf{r},t\right)\iint\mbox{\ensuremath{\int}d\textbf{r}}'d\mathbf{r}''d\mathbf{r}'''\nonumber \\
v_{C}\left(\left|{\bf r}-\mathbf{r}''\right|\right) & \tilde{\chi}_{RPA}\left({\bf \mathbf{r}}'',{\bf r}''',0\right)\left(v_{C}\left(\left|{\bf r}'''-{\bf r}'\right|\right)\right.\nonumber \\
+ & \left.f_{XC}\mbox{\ensuremath{\left(\mathbf{r}'''\right)}}\delta\mbox{\ensuremath{\left({\bf r}'''-{\bf r}'\right)}}\right)\eta_{x}^{\gamma}\left(\mathbf{r}',t\right)^{*}\psi\left(\mathbf{r}',t\right).\label{eq:deltaKex}
\end{align}
Define a perturbation potential 
\begin{align}
v^{\gamma}\left(\mathbf{r},t\right) & =\int d\mathbf{r}'\left(v_{C}\left(\left|{\bf r}'''-{\bf r}'\right|\right)\right.\nonumber \\
+ & \left.f_{XC}\mbox{\ensuremath{\left(\mathbf{r}'''\right)}}\delta\mbox{\ensuremath{\left({\bf r}'''-{\bf r}'\right)}}\right)\eta_{x}^{\gamma}\left(\mathbf{r}',t\right)^{*}\psi\left(\mathbf{r}',t\right)
\end{align}
 and rewrite Eq.~\eqref{eq:deltaKex} as:
\begin{align}
\delta\hat{k}_{\epsilon X}^{\gamma}\left(t\right)\psi\left(\mathbf{r},t\right) & =\eta_{x}^{\gamma}\left(\mathbf{r},t\right)\iint d\mathbf{r}'d\mathbf{r}''\nonumber \\
v_{C}\left(\left|{\bf r}-\mathbf{r}'\right|\right) & \tilde{\chi}_{RPA}\left({\bf \mathbf{r}}',{\bf r}'',0\right)v^{\gamma}\left(\mathbf{r}'',t\right).
\end{align}
The action of $\tilde{\chi}_{RPA}\left({\bf \mathbf{r}}',{\bf r}'',0\right)$
on $v^{\gamma}\left(\mathbf{r}'',t\right)$ is manageable by using
a stochastic TDDFT algorithm:\cite{Gao2014TDsDFT}
\begin{enumerate}
\item Take$N_{RPA}$ projected stochastic orbitals from the $N_{\zeta}$
generated above. If $N_{RPA}>N_{\zeta}$ generate additional projected
stochastic orbitals following the prescription given in 1 and 2 above.
This needs to be done just once, i.e., at the beginning of the calculation,
generate enough projected stochastic orbitals to be used throughout
the calculation. 
\item Apply a perturbation at $\tau=0$: $\chi_{j}^{\gamma'}\left({\bf r},\tau=0\right)=e^{-i\gamma'v^{\gamma}\left({\bf r},t\right)/\hbar}\xi_{j}\left({\bf r}\right)$,
where $\gamma'$ is the strength of the RPA perturbation. Note that
at each time $t$ used for solving the TD\textit{s}BSE, one has to
apply a different perturbation $v^{\gamma}\left({\bf r},t\right)$
at $\tau=0$, which is used to indicate the time for the RPA propagation.
\item Propagate the orbitals using the adiabatic stochastic time-dependent
equations:
\begin{align}
i\hbar\frac{\partial\chi_{j}^{\gamma'}\left({\bf r},\tau\right)}{\partial\tau} & =\hat{h}_{RPA}^{\gamma'}\left(\tau\right)\chi_{j}^{\gamma'}\left({\bf r},\tau\right),\label{eq:tdsdft-1}
\end{align}
Here, one can take $\hat{h}_{RPA}^{\gamma'}\left(\tau\right)=\hat{h}_{qp}$
or $\hat{h}_{RPA}^{\gamma'}\left(\tau\right)=\hat{h}_{qp}+v_{HXC}\left[n_{RPA}^{\gamma'}\left(\tau\right)\right]\left(\mathbf{r}\right)-v_{HXC}\left[n_{RPA}^{0}\left(\tau\right)\right]\left(\mathbf{r}\right)$.
For the latter case, $v_{HXC}\left[n\right]\left(\mathbf{r}\right)=\int d\mathbf{r}'\frac{n\left(\mathbf{r}'\right)}{\left|\mathbf{r}-\mathbf{r}'\right|}+v_{XC}\left(n\left(\mathbf{r}\right)\right)$
and $v_{XC}\left(n\left(\mathbf{r}\right)\right)$ is the local density
(or semi-local) approximation for the exchange correlation potential.
The density is obtained as an \emph{average }over the RPA stochastic
orbital densities: 
\begin{align}
n_{RPA}^{\gamma'}\left({\bf {\bf r}},\tau\right) & =\frac{2}{N_{RPA}}\sum_{j=1}^{N_{RPA}}\left|\chi_{j}^{\gamma'}\left({\bf r},\tau\right)\right|^{2}\label{eq:dens-from-xi-2}
\end{align}

\item Generate $\Delta n_{RPA}\left({\bf r},\tau\right)=\frac{1}{\gamma'}\left(n_{RPA}^{\gamma'}\left({\bf r},\tau\right)-n_{RPA}^{0}\left({\bf r},\tau\right)\right)$
and its half Fourier transformed quantity $\Delta\tilde{n}_{RPA}\left(\mathbf{r},0\right)$
at $\omega=0$.
\item Obtain the action of $\delta\hat{k}_{\epsilon X}^{\gamma}\left(t\right)\psi\left(\mathbf{r},t\right)=\eta_{x}^{\gamma}\left(\mathbf{r},t\right)\iint d\mathbf{r}'d\mathbf{r}''v_{C}\left(\left|{\bf r}-\mathbf{r}'\right|\right)\tilde{\chi}_{RPA}\left({\bf \mathbf{r}}',{\bf r}'',0\right)v^{\gamma}\left(\mathbf{r}'',t\right)$
from $\delta\hat{k}_{\epsilon X}\left(t\right)\psi\left(\mathbf{r},t\right)=\eta_{x}^{\gamma}\left(\mathbf{r},t\right)\iint d\mathbf{r}'d\mathbf{r}''v_{C}\left(\left|{\bf r}-\mathbf{r}'\right|\right)\Delta\tilde{n}_{RPA}\left({\bf r}',0\right)$.
\end{enumerate}
Step 1-5 need to be repeated at each time step $\Delta t$ of the
TD\emph{s}BSE propagation.

\subsection*{Appendix B: Ehrenfest theorem}

Ehrenfest theorem asserts that a correct propagation must preserve
the relation
\begin{equation}
\left\langle \dot{\mathbf{q}}\left(t\right)\right\rangle =i\left\langle \left[\hat{h}_{BS},\hat{\mathbf{q}}\right]\right\rangle 
\end{equation}
For a TDBSE this relation is given by

\begin{equation}
i\left\langle \left[\hat{h}_{BS},\hat{\mathbf{q}}\right]\right\rangle =\frac{\left\langle \mathbf{p}\left(t\right)\right\rangle }{m}+i\left\langle \left[\hat{k}_{\epsilon X}^{\gamma}\left(t\right)-\hat{k}_{\epsilon X}^{0}\left(t\right),\hat{\mathbf{q}}\right]\right\rangle 
\end{equation}
where $\hat{k}_{\epsilon X}^{\gamma}\left(t\right)\psi\left(\mathbf{r}\right)=-\frac{1}{2}\int d\mathbf{r}'\rho^{\gamma}\left(\mathbf{r},{\bf r}',t\right)W^{RPA}\left({\bf r},{\bf r}',0\right)\psi\left(\mathbf{r}'\right)$.
To satisfy the Ehrenfest theorem $\left\langle \left[\hat{k}_{\epsilon X}^{\gamma}\left(t\right)-\hat{k}_{\epsilon X}^{0}\left(t\right),\hat{\mathbf{q}}\right]\right\rangle $
should vanish. The commutator of the exchange operator is given by:

\begin{eqnarray}
i\left\langle \left[\hat{k}_{\epsilon X}^{\gamma}\left(t\right),\hat{\mathbf{q}}\right]\right\rangle  & = & -\frac{i}{2}\iint d^{3}rd^{3}r'\left|\rho^{\gamma}\left(\mathbf{r},{\bf r}',t\right)\right|^{2}\nonumber \\
 & \times & W^{RPA}\left({\bf r},{\bf r}',0\right)\left(\mathbf{r}-\mathbf{r}'\right).
\end{eqnarray}
In the above, the commuter vanishes for the $\hat{k}_{\epsilon X}^{\gamma}\left(t\right)$
term due to symmetry, but there is no \textit{a-priori} reason why
the $\hat{k}_{\epsilon X}^{0}\left(t\right)$ term should vanish.
However, as illustrated numerically in Fig.~\ref{fig:Average-momentum-calculated},
the contribution of this non-vanishing term is rather small even on
timescales much larger than the typical frequency in the system.


\begin{thebibliography}{125}%
\makeatletter
\providecommand \@ifxundefined [1]{%
 \@ifx{#1\undefined}
}%
\providecommand \@ifnum [1]{%
 \ifnum #1\expandafter \@firstoftwo
 \else \expandafter \@secondoftwo
 \fi
}%
\providecommand \@ifx [1]{%
 \ifx #1\expandafter \@firstoftwo
 \else \expandafter \@secondoftwo
 \fi
}%
\providecommand \natexlab [1]{#1}%
\providecommand \enquote  [1]{``#1''}%
\providecommand \bibnamefont  [1]{#1}%
\providecommand \bibfnamefont [1]{#1}%
\providecommand \citenamefont [1]{#1}%
\providecommand \href@noop [0]{\@secondoftwo}%
\providecommand \href [0]{\begingroup \@sanitize@url \@href}%
\providecommand \@href[1]{\@@startlink{#1}\@@href}%
\providecommand \@@href[1]{\endgroup#1\@@endlink}%
\providecommand \@sanitize@url [0]{\catcode `\\12\catcode `\$12\catcode
  `\&12\catcode `\#12\catcode `\^12\catcode `\_12\catcode `\%12\relax}%
\providecommand \@@startlink[1]{}%
\providecommand \@@endlink[0]{}%
\providecommand \url  [0]{\begingroup\@sanitize@url \@url }%
\providecommand \@url [1]{\endgroup\@href {#1}{\urlprefix }}%
\providecommand \urlprefix  [0]{URL }%
\providecommand \Eprint [0]{\href }%
\providecommand \doibase [0]{http://dx.doi.org/}%
\providecommand \selectlanguage [0]{\@gobble}%
\providecommand \bibinfo  [0]{\@secondoftwo}%
\providecommand \bibfield  [0]{\@secondoftwo}%
\providecommand \translation [1]{[#1]}%
\providecommand \BibitemOpen [0]{}%
\providecommand \bibitemStop [0]{}%
\providecommand \bibitemNoStop [0]{.\EOS\space}%
\providecommand \EOS [0]{\spacefactor3000\relax}%
\providecommand \BibitemShut  [1]{\csname bibitem#1\endcsname}%
\let\auto@bib@innerbib\@empty
\bibitem [{\citenamefont {Coe}, \citenamefont {Woo},\ and\ \citenamefont
  {Moungi~Bawendi}(2002)}]{Coe2002a}%
  \BibitemOpen
  \bibfield  {author} {\bibinfo {author} {\bibfnamefont {S.}~\bibnamefont
  {Coe}}, \bibinfo {author} {\bibfnamefont {W.-K.}\ \bibnamefont {Woo}}, \ and\
  \bibinfo {author} {\bibfnamefont {V.~B.}\ \bibnamefont {Moungi~Bawendi}},\
  }\href@noop {} {\bibfield  {journal} {\bibinfo  {journal} {Nature}\ }\textbf
  {\bibinfo {volume} {420}},\ \bibinfo {pages} {800} (\bibinfo {year}
  {2002})}\BibitemShut {NoStop}%
\bibitem [{\citenamefont {Tessler}\ \emph {et~al.}(2002)\citenamefont
  {Tessler}, \citenamefont {Medvedev}, \citenamefont {Kazes}, \citenamefont
  {Kan},\ and\ \citenamefont {Banin}}]{Tessler2002}%
  \BibitemOpen
  \bibfield  {author} {\bibinfo {author} {\bibfnamefont {N.}~\bibnamefont
  {Tessler}}, \bibinfo {author} {\bibfnamefont {V.}~\bibnamefont {Medvedev}},
  \bibinfo {author} {\bibfnamefont {M.}~\bibnamefont {Kazes}}, \bibinfo
  {author} {\bibfnamefont {S.~H.}\ \bibnamefont {Kan}}, \ and\ \bibinfo
  {author} {\bibfnamefont {U.}~\bibnamefont {Banin}},\ }\href@noop {}
  {\bibfield  {journal} {\bibinfo  {journal} {Science}\ }\textbf {\bibinfo
  {volume} {295}},\ \bibinfo {pages} {1506} (\bibinfo {year}
  {2002})}\BibitemShut {NoStop}%
\bibitem [{\citenamefont {Gur}\ \emph {et~al.}(2005)\citenamefont {Gur},
  \citenamefont {Fromer}, \citenamefont {Geier},\ and\ \citenamefont
  {Alivisatos}}]{Gur2005}%
  \BibitemOpen
  \bibfield  {author} {\bibinfo {author} {\bibfnamefont {I.}~\bibnamefont
  {Gur}}, \bibinfo {author} {\bibfnamefont {N.~A.}\ \bibnamefont {Fromer}},
  \bibinfo {author} {\bibfnamefont {M.~L.}\ \bibnamefont {Geier}}, \ and\
  \bibinfo {author} {\bibfnamefont {A.~P.}\ \bibnamefont {Alivisatos}},\
  }\href@noop {} {\bibfield  {journal} {\bibinfo  {journal} {Science}\ }\textbf
  {\bibinfo {volume} {310}},\ \bibinfo {pages} {462} (\bibinfo {year}
  {2005})}\BibitemShut {NoStop}%
\bibitem [{\citenamefont {Talapin}\ and\ \citenamefont
  {Murray}(2005)}]{Talapin2005}%
  \BibitemOpen
  \bibfield  {author} {\bibinfo {author} {\bibfnamefont {D.~V.}\ \bibnamefont
  {Talapin}}\ and\ \bibinfo {author} {\bibfnamefont {C.~B.}\ \bibnamefont
  {Murray}},\ }\href@noop {} {\bibfield  {journal} {\bibinfo  {journal}
  {Science}\ }\textbf {\bibinfo {volume} {310}},\ \bibinfo {pages} {86}
  (\bibinfo {year} {2005})}\BibitemShut {NoStop}%
\bibitem [{\citenamefont {Runge}\ and\ \citenamefont
  {Gross}(1984)}]{Runge1984}%
  \BibitemOpen
  \bibfield  {author} {\bibinfo {author} {\bibfnamefont {E.}~\bibnamefont
  {Runge}}\ and\ \bibinfo {author} {\bibfnamefont {E.~K.~U.}\ \bibnamefont
  {Gross}},\ }\href@noop {} {\bibfield  {journal} {\bibinfo  {journal} {Phys.
  Rev. Lett.}\ }\textbf {\bibinfo {volume} {52}},\ \bibinfo {pages} {997}
  (\bibinfo {year} {1984})}\BibitemShut {NoStop}%
\bibitem [{\citenamefont {van Leeuwen}(2001)}]{Leeuwen2001}%
  \BibitemOpen
  \bibfield  {author} {\bibinfo {author} {\bibfnamefont {R.}~\bibnamefont {van
  Leeuwen}},\ }\href@noop {} {\bibfield  {journal} {\bibinfo  {journal} {Inter.
  J. Moder. Phys. B}\ }\textbf {\bibinfo {volume} {15}},\ \bibinfo {pages}
  {1969} (\bibinfo {year} {2001})}\BibitemShut {NoStop}%
\bibitem [{\citenamefont {Onida}, \citenamefont {Reining},\ and\ \citenamefont
  {Rubio}(2002)}]{Onida2002}%
  \BibitemOpen
  \bibfield  {author} {\bibinfo {author} {\bibfnamefont {G.}~\bibnamefont
  {Onida}}, \bibinfo {author} {\bibfnamefont {L.}~\bibnamefont {Reining}}, \
  and\ \bibinfo {author} {\bibfnamefont {A.}~\bibnamefont {Rubio}},\
  }\href@noop {} {\bibfield  {journal} {\bibinfo  {journal} {Rev. Mod. Phys.}\
  }\textbf {\bibinfo {volume} {74}},\ \bibinfo {pages} {601} (\bibinfo {year}
  {2002})}\BibitemShut {NoStop}%
\bibitem [{\citenamefont {Maitra}\ \emph {et~al.}(2002)\citenamefont {Maitra},
  \citenamefont {Burke}, \citenamefont {Appel},\ and\ \citenamefont
  {Gross}}]{Maitra2002a}%
  \BibitemOpen
  \bibfield  {author} {\bibinfo {author} {\bibfnamefont {N.~T.}\ \bibnamefont
  {Maitra}}, \bibinfo {author} {\bibfnamefont {K.}~\bibnamefont {Burke}},
  \bibinfo {author} {\bibfnamefont {H.}~\bibnamefont {Appel}}, \ and\ \bibinfo
  {author} {\bibfnamefont {E.~K.~U.}\ \bibnamefont {Gross}},\ }\enquote
  {\bibinfo {title} {Ten topical questions in time dependent density functional
  theory},}\ in\ \href@noop {} {\emph {\bibinfo {booktitle} {Reviews in Modern
  Quantum Chemistry: A celebration of the contributions of R. G. Parr}}},\
  Vol.~\bibinfo {volume} {II},\ \bibinfo {editor} {edited by\ \bibinfo {editor}
  {\bibfnamefont {K.~D.}\ \bibnamefont {Sen}}}\ (\bibinfo  {publisher}
  {World-Scientific},\ \bibinfo {address} {Singapore},\ \bibinfo {year}
  {2002})\ p.\ \bibinfo {pages} {1186}\BibitemShut {NoStop}%
\bibitem [{\citenamefont {Marques}\ and\ \citenamefont
  {Gross}(2004)}]{Marques2004}%
  \BibitemOpen
  \bibfield  {author} {\bibinfo {author} {\bibfnamefont {M.}~\bibnamefont
  {Marques}}\ and\ \bibinfo {author} {\bibfnamefont {E.}~\bibnamefont
  {Gross}},\ }\href@noop {} {\bibfield  {journal} {\bibinfo  {journal} {Annu.
  Rev. Phys. Chem.}\ }\textbf {\bibinfo {volume} {55}},\ \bibinfo {pages} {427}
  (\bibinfo {year} {2004})}\BibitemShut {NoStop}%
\bibitem [{\citenamefont {Burke}, \citenamefont {Werschnik},\ and\
  \citenamefont {Gross}(2005)}]{Burke2005a}%
  \BibitemOpen
  \bibfield  {author} {\bibinfo {author} {\bibfnamefont {K.}~\bibnamefont
  {Burke}}, \bibinfo {author} {\bibfnamefont {J.}~\bibnamefont {Werschnik}}, \
  and\ \bibinfo {author} {\bibfnamefont {E.~K.~U.}\ \bibnamefont {Gross}},\
  }\href@noop {} {\bibfield  {journal} {\bibinfo  {journal} {J. Chem. Phys.}\
  }\textbf {\bibinfo {volume} {123}},\ \bibinfo {pages} {062206} (\bibinfo
  {year} {2005})}\BibitemShut {NoStop}%
\bibitem [{\citenamefont {Botti}\ \emph {et~al.}(2007)\citenamefont {Botti},
  \citenamefont {Schindlmayr}, \citenamefont {Del~Sole},\ and\ \citenamefont
  {Reining}}]{Botti2007}%
  \BibitemOpen
  \bibfield  {author} {\bibinfo {author} {\bibfnamefont {S.}~\bibnamefont
  {Botti}}, \bibinfo {author} {\bibfnamefont {A.}~\bibnamefont {Schindlmayr}},
  \bibinfo {author} {\bibfnamefont {R.}~\bibnamefont {Del~Sole}}, \ and\
  \bibinfo {author} {\bibfnamefont {L.}~\bibnamefont {Reining}},\ }\href@noop
  {} {\bibfield  {journal} {\bibinfo  {journal} {Rep. Prog. Phys.}\ }\textbf
  {\bibinfo {volume} {70}},\ \bibinfo {pages} {357} (\bibinfo {year}
  {2007})}\BibitemShut {NoStop}%
\bibitem [{\citenamefont {Jacquemin}\ \emph {et~al.}(2009)\citenamefont
  {Jacquemin}, \citenamefont {Perpete}, \citenamefont {Ciofini},\ and\
  \citenamefont {Adamo}}]{Jacquemin2009}%
  \BibitemOpen
  \bibfield  {author} {\bibinfo {author} {\bibfnamefont {D.}~\bibnamefont
  {Jacquemin}}, \bibinfo {author} {\bibfnamefont {E.~A.}\ \bibnamefont
  {Perpete}}, \bibinfo {author} {\bibfnamefont {I.}~\bibnamefont {Ciofini}}, \
  and\ \bibinfo {author} {\bibfnamefont {C.}~\bibnamefont {Adamo}},\
  }\href@noop {} {\bibfield  {journal} {\bibinfo  {journal} {Acc. Chem. Res.}\
  }\textbf {\bibinfo {volume} {42}},\ \bibinfo {pages} {326} (\bibinfo {year}
  {2009})}\BibitemShut {NoStop}%
\bibitem [{\citenamefont {Casida}(2009)}]{Casida2009}%
  \BibitemOpen
  \bibfield  {author} {\bibinfo {author} {\bibfnamefont {M.~E.}\ \bibnamefont
  {Casida}},\ }\href@noop {} {\bibfield  {journal} {\bibinfo  {journal} {J.
  Mol. Struct.}\ }\textbf {\bibinfo {volume} {914}},\ \bibinfo {pages} {3}
  (\bibinfo {year} {2009})}\BibitemShut {NoStop}%
\bibitem [{\citenamefont {Adamo}\ and\ \citenamefont
  {Jacquemin}(2013)}]{Adamo2013}%
  \BibitemOpen
  \bibfield  {author} {\bibinfo {author} {\bibfnamefont {C.}~\bibnamefont
  {Adamo}}\ and\ \bibinfo {author} {\bibfnamefont {D.}~\bibnamefont
  {Jacquemin}},\ }\href@noop {} {\bibfield  {journal} {\bibinfo  {journal}
  {Chem. Soc. Rev.}\ }\textbf {\bibinfo {volume} {42}},\ \bibinfo {pages} {845}
  (\bibinfo {year} {2013})}\BibitemShut {NoStop}%
\bibitem [{\citenamefont {Stratmann}, \citenamefont {Scuseria},\ and\
  \citenamefont {Frisch}(1998)}]{Stratmann1998}%
  \BibitemOpen
  \bibfield  {author} {\bibinfo {author} {\bibfnamefont {R.~E.}\ \bibnamefont
  {Stratmann}}, \bibinfo {author} {\bibfnamefont {G.~E.}\ \bibnamefont
  {Scuseria}}, \ and\ \bibinfo {author} {\bibfnamefont {M.~J.}\ \bibnamefont
  {Frisch}},\ }\href@noop {} {\bibfield  {journal} {\bibinfo  {journal} {J.
  Chem. Phys.}\ }\textbf {\bibinfo {volume} {109}},\ \bibinfo {pages} {8218}
  (\bibinfo {year} {1998})}\BibitemShut {NoStop}%
\bibitem [{\citenamefont {Yabana}\ and\ \citenamefont
  {Bertsch}(1996)}]{Yabana1996}%
  \BibitemOpen
  \bibfield  {author} {\bibinfo {author} {\bibfnamefont {K.}~\bibnamefont
  {Yabana}}\ and\ \bibinfo {author} {\bibfnamefont {G.~F.}\ \bibnamefont
  {Bertsch}},\ }\href@noop {} {\bibfield  {journal} {\bibinfo  {journal} {Phys.
  Rev. B}\ }\textbf {\bibinfo {volume} {54}},\ \bibinfo {pages} {4484}
  (\bibinfo {year} {1996})}\BibitemShut {NoStop}%
\bibitem [{\citenamefont {Bertsch}\ \emph {et~al.}(2000)\citenamefont
  {Bertsch}, \citenamefont {Iwata}, \citenamefont {Rubio},\ and\ \citenamefont
  {Yabana}}]{Bertsch2000}%
  \BibitemOpen
  \bibfield  {author} {\bibinfo {author} {\bibfnamefont {G.~F.}\ \bibnamefont
  {Bertsch}}, \bibinfo {author} {\bibfnamefont {J.~I.}\ \bibnamefont {Iwata}},
  \bibinfo {author} {\bibfnamefont {A.}~\bibnamefont {Rubio}}, \ and\ \bibinfo
  {author} {\bibfnamefont {K.}~\bibnamefont {Yabana}},\ }\href@noop {}
  {\bibfield  {journal} {\bibinfo  {journal} {Phys. Rev. B}\ }\textbf {\bibinfo
  {volume} {62}},\ \bibinfo {pages} {7998} (\bibinfo {year}
  {2000})}\BibitemShut {NoStop}%
\bibitem [{\citenamefont {Baer}\ and\ \citenamefont
  {Neuhauser}(2004)}]{Baer2004b}%
  \BibitemOpen
  \bibfield  {author} {\bibinfo {author} {\bibfnamefont {R.}~\bibnamefont
  {Baer}}\ and\ \bibinfo {author} {\bibfnamefont {D.}~\bibnamefont
  {Neuhauser}},\ }\href@noop {} {\bibfield  {journal} {\bibinfo  {journal} {J.
  Chem. Phys.}\ }\textbf {\bibinfo {volume} {121}},\ \bibinfo {pages} {9803}
  (\bibinfo {year} {2004})}\BibitemShut {NoStop}%
\bibitem [{\citenamefont {Bauernschmitt}\ and\ \citenamefont
  {Ahlrichs}(1996)}]{Bauernschmitt1996}%
  \BibitemOpen
  \bibfield  {author} {\bibinfo {author} {\bibfnamefont {R.}~\bibnamefont
  {Bauernschmitt}}\ and\ \bibinfo {author} {\bibfnamefont {R.}~\bibnamefont
  {Ahlrichs}},\ }\href@noop {} {\bibfield  {journal} {\bibinfo  {journal}
  {Chem. Phys. Lett.}\ }\textbf {\bibinfo {volume} {256}},\ \bibinfo {pages}
  {454} (\bibinfo {year} {1996})}\BibitemShut {NoStop}%
\bibitem [{\citenamefont {Bauernschmitt}\ \emph {et~al.}(1998)\citenamefont
  {Bauernschmitt}, \citenamefont {Ahlrichs}, \citenamefont {Hennrich},\ and\
  \citenamefont {Kappes}}]{Bauernschmitt1998}%
  \BibitemOpen
  \bibfield  {author} {\bibinfo {author} {\bibfnamefont {R.}~\bibnamefont
  {Bauernschmitt}}, \bibinfo {author} {\bibfnamefont {R.}~\bibnamefont
  {Ahlrichs}}, \bibinfo {author} {\bibfnamefont {F.~H.}\ \bibnamefont
  {Hennrich}}, \ and\ \bibinfo {author} {\bibfnamefont {M.~M.}\ \bibnamefont
  {Kappes}},\ }\href@noop {} {\bibfield  {journal} {\bibinfo  {journal} {J. Am.
  Chem. Soc.}\ }\textbf {\bibinfo {volume} {120}},\ \bibinfo {pages} {5052}
  (\bibinfo {year} {1998})}\BibitemShut {NoStop}%
\bibitem [{\citenamefont {Fabian}(2001)}]{Fabian2001}%
  \BibitemOpen
  \bibfield  {author} {\bibinfo {author} {\bibfnamefont {J.}~\bibnamefont
  {Fabian}},\ }\href@noop {} {\bibfield  {journal} {\bibinfo  {journal} {Theor.
  Chem. Acc.}\ }\textbf {\bibinfo {volume} {106}},\ \bibinfo {pages} {199}
  (\bibinfo {year} {2001})}\BibitemShut {NoStop}%
\bibitem [{\citenamefont {Vasiliev}, \citenamefont {Ogut},\ and\ \citenamefont
  {Chelikowsky}(2002)}]{Vasiliev2002}%
  \BibitemOpen
  \bibfield  {author} {\bibinfo {author} {\bibfnamefont {I.}~\bibnamefont
  {Vasiliev}}, \bibinfo {author} {\bibfnamefont {S.}~\bibnamefont {Ogut}}, \
  and\ \bibinfo {author} {\bibfnamefont {J.~R.}\ \bibnamefont {Chelikowsky}},\
  }\href@noop {} {\bibfield  {journal} {\bibinfo  {journal} {Phys. Rev. B}\
  }\textbf {\bibinfo {volume} {65}},\ \bibinfo {pages} {115416} (\bibinfo
  {year} {2002})}\BibitemShut {NoStop}%
\bibitem [{\citenamefont {Shao}, \citenamefont {Head-Gordon},\ and\
  \citenamefont {Krylov}(2003)}]{Shao2003}%
  \BibitemOpen
  \bibfield  {author} {\bibinfo {author} {\bibfnamefont {Y.~H.}\ \bibnamefont
  {Shao}}, \bibinfo {author} {\bibfnamefont {M.}~\bibnamefont {Head-Gordon}}, \
  and\ \bibinfo {author} {\bibfnamefont {A.~I.}\ \bibnamefont {Krylov}},\
  }\href@noop {} {\bibfield  {journal} {\bibinfo  {journal} {J. Chem. Phys.}\
  }\textbf {\bibinfo {volume} {118}},\ \bibinfo {pages} {4807} (\bibinfo {year}
  {2003})}\BibitemShut {NoStop}%
\bibitem [{\citenamefont {Troparevsky}, \citenamefont {Kronik},\ and\
  \citenamefont {Chelikowsky}(2003)}]{Troparevsky2003}%
  \BibitemOpen
  \bibfield  {author} {\bibinfo {author} {\bibfnamefont {M.~C.}\ \bibnamefont
  {Troparevsky}}, \bibinfo {author} {\bibfnamefont {L.}~\bibnamefont {Kronik}},
  \ and\ \bibinfo {author} {\bibfnamefont {J.~R.}\ \bibnamefont
  {Chelikowsky}},\ }\href@noop {} {\bibfield  {journal} {\bibinfo  {journal}
  {J. Chem. Phys.}\ }\textbf {\bibinfo {volume} {119}},\ \bibinfo {pages}
  {2284} (\bibinfo {year} {2003})}\BibitemShut {NoStop}%
\bibitem [{\citenamefont {Maitra}(2005)}]{Maitra2005}%
  \BibitemOpen
  \bibfield  {author} {\bibinfo {author} {\bibfnamefont {N.~T.}\ \bibnamefont
  {Maitra}},\ }\href@noop {} {\bibfield  {journal} {\bibinfo  {journal} {J.
  Chem. Phys.}\ }\textbf {\bibinfo {volume} {122}},\ \bibinfo {pages} {234104}
  (\bibinfo {year} {2005})}\BibitemShut {NoStop}%
\bibitem [{\citenamefont {Andzelm}\ \emph {et~al.}(2007)\citenamefont
  {Andzelm}, \citenamefont {Rawlett}, \citenamefont {Orlicki},\ and\
  \citenamefont {Snyder}}]{Andzelm2007}%
  \BibitemOpen
  \bibfield  {author} {\bibinfo {author} {\bibfnamefont {J.}~\bibnamefont
  {Andzelm}}, \bibinfo {author} {\bibfnamefont {A.~M.}\ \bibnamefont
  {Rawlett}}, \bibinfo {author} {\bibfnamefont {J.~A.}\ \bibnamefont
  {Orlicki}}, \ and\ \bibinfo {author} {\bibfnamefont {J.~F.}\ \bibnamefont
  {Snyder}},\ }\href@noop {} {\bibfield  {journal} {\bibinfo  {journal} {J.
  Chem. Theory Comput.}\ }\textbf {\bibinfo {volume} {3}},\ \bibinfo {pages}
  {870} (\bibinfo {year} {2007})}\BibitemShut {NoStop}%
\bibitem [{\citenamefont {Govind}\ \emph {et~al.}(2009)\citenamefont {Govind},
  \citenamefont {Valiev}, \citenamefont {Jensen},\ and\ \citenamefont
  {Kowalski}}]{Govind2009}%
  \BibitemOpen
  \bibfield  {author} {\bibinfo {author} {\bibfnamefont {N.}~\bibnamefont
  {Govind}}, \bibinfo {author} {\bibfnamefont {M.}~\bibnamefont {Valiev}},
  \bibinfo {author} {\bibfnamefont {L.}~\bibnamefont {Jensen}}, \ and\ \bibinfo
  {author} {\bibfnamefont {K.}~\bibnamefont {Kowalski}},\ }\href@noop {}
  {\bibfield  {journal} {\bibinfo  {journal} {J. Phys. Chem. A}\ }\textbf
  {\bibinfo {volume} {113}},\ \bibinfo {pages} {6041} (\bibinfo {year}
  {2009})}\BibitemShut {NoStop}%
\bibitem [{\citenamefont {Peach}\ \emph {et~al.}(2009)\citenamefont {Peach},
  \citenamefont {Le~Sueur}, \citenamefont {Ruud}, \citenamefont {Guillaume},\
  and\ \citenamefont {Tozer}}]{Peach2009}%
  \BibitemOpen
  \bibfield  {author} {\bibinfo {author} {\bibfnamefont {M.~J.~G.}\
  \bibnamefont {Peach}}, \bibinfo {author} {\bibfnamefont {C.~R.}\ \bibnamefont
  {Le~Sueur}}, \bibinfo {author} {\bibfnamefont {K.}~\bibnamefont {Ruud}},
  \bibinfo {author} {\bibfnamefont {M.}~\bibnamefont {Guillaume}}, \ and\
  \bibinfo {author} {\bibfnamefont {D.~J.}\ \bibnamefont {Tozer}},\ }\href@noop
  {} {\bibfield  {journal} {\bibinfo  {journal} {Phys. Chem. Chem. Phys.}\
  }\textbf {\bibinfo {volume} {11}},\ \bibinfo {pages} {4465} (\bibinfo {year}
  {2009})}\BibitemShut {NoStop}%
\bibitem [{\citenamefont {Kuritz}\ \emph {et~al.}(2011)\citenamefont {Kuritz},
  \citenamefont {Stein}, \citenamefont {Baer},\ and\ \citenamefont
  {Kronik}}]{Kuritz2011}%
  \BibitemOpen
  \bibfield  {author} {\bibinfo {author} {\bibfnamefont {N.}~\bibnamefont
  {Kuritz}}, \bibinfo {author} {\bibfnamefont {T.}~\bibnamefont {Stein}},
  \bibinfo {author} {\bibfnamefont {R.}~\bibnamefont {Baer}}, \ and\ \bibinfo
  {author} {\bibfnamefont {L.}~\bibnamefont {Kronik}},\ }\href@noop {}
  {\bibfield  {journal} {\bibinfo  {journal} {J. Chem. Theory Comput.}\
  }\textbf {\bibinfo {volume} {7}},\ \bibinfo {pages} {2408} (\bibinfo {year}
  {2011})}\BibitemShut {NoStop}%
\bibitem [{\citenamefont {Peach}, \citenamefont {Williamson},\ and\
  \citenamefont {Tozer}(2011)}]{Peach2011}%
  \BibitemOpen
  \bibfield  {author} {\bibinfo {author} {\bibfnamefont {M.~J.~G.}\
  \bibnamefont {Peach}}, \bibinfo {author} {\bibfnamefont {M.~J.}\ \bibnamefont
  {Williamson}}, \ and\ \bibinfo {author} {\bibfnamefont {D.~J.}\ \bibnamefont
  {Tozer}},\ }\href@noop {} {\bibfield  {journal} {\bibinfo  {journal} {J.
  Chem. Theory Comput.}\ }\textbf {\bibinfo {volume} {7}},\ \bibinfo {pages}
  {3578} (\bibinfo {year} {2011})}\BibitemShut {NoStop}%
\bibitem [{\citenamefont {Srebro}\ \emph {et~al.}(2011)\citenamefont {Srebro},
  \citenamefont {Govind}, \citenamefont {de~Jong},\ and\ \citenamefont
  {Autschbach}}]{Srebro2011a}%
  \BibitemOpen
  \bibfield  {author} {\bibinfo {author} {\bibfnamefont {M.}~\bibnamefont
  {Srebro}}, \bibinfo {author} {\bibfnamefont {N.}~\bibnamefont {Govind}},
  \bibinfo {author} {\bibfnamefont {W.~A.}\ \bibnamefont {de~Jong}}, \ and\
  \bibinfo {author} {\bibfnamefont {J.}~\bibnamefont {Autschbach}},\
  }\href@noop {} {\bibfield  {journal} {\bibinfo  {journal} {J. Phys. Chem. A}\
  }\textbf {\bibinfo {volume} {115}},\ \bibinfo {pages} {10930} (\bibinfo
  {year} {2011})}\BibitemShut {NoStop}%
\bibitem [{\citenamefont {Richard}\ and\ \citenamefont
  {Herbert}(2011)}]{Richard2011}%
  \BibitemOpen
  \bibfield  {author} {\bibinfo {author} {\bibfnamefont {R.~M.}\ \bibnamefont
  {Richard}}\ and\ \bibinfo {author} {\bibfnamefont {J.~M.}\ \bibnamefont
  {Herbert}},\ }\href@noop {} {\bibfield  {journal} {\bibinfo  {journal} {J.
  Chem. Theory Comput.}\ }\textbf {\bibinfo {volume} {7}},\ \bibinfo {pages}
  {1296} (\bibinfo {year} {2011})}\BibitemShut {NoStop}%
\bibitem [{\citenamefont {Chantzis}\ \emph {et~al.}(2013)\citenamefont
  {Chantzis}, \citenamefont {Laurent}, \citenamefont {Adamo},\ and\
  \citenamefont {Jacquemin}}]{Chantzis2013}%
  \BibitemOpen
  \bibfield  {author} {\bibinfo {author} {\bibfnamefont {A.}~\bibnamefont
  {Chantzis}}, \bibinfo {author} {\bibfnamefont {A.~D.}\ \bibnamefont
  {Laurent}}, \bibinfo {author} {\bibfnamefont {C.}~\bibnamefont {Adamo}}, \
  and\ \bibinfo {author} {\bibfnamefont {D.}~\bibnamefont {Jacquemin}},\
  }\href@noop {} {\bibfield  {journal} {\bibinfo  {journal} {J. Chem. Theory
  Comput.}\ }\textbf {\bibinfo {volume} {9}},\ \bibinfo {pages} {4517}
  (\bibinfo {year} {2013})}\BibitemShut {NoStop}%
\bibitem [{\citenamefont {Bauernschmitt}\ \emph {et~al.}(1997)\citenamefont
  {Bauernschmitt}, \citenamefont {Haser}, \citenamefont {Treutler},\ and\
  \citenamefont {Ahlrichs}}]{Bauernschmitt1997}%
  \BibitemOpen
  \bibfield  {author} {\bibinfo {author} {\bibfnamefont {R.}~\bibnamefont
  {Bauernschmitt}}, \bibinfo {author} {\bibfnamefont {M.}~\bibnamefont
  {Haser}}, \bibinfo {author} {\bibfnamefont {O.}~\bibnamefont {Treutler}}, \
  and\ \bibinfo {author} {\bibfnamefont {R.}~\bibnamefont {Ahlrichs}},\
  }\href@noop {} {\bibfield  {journal} {\bibinfo  {journal} {Chem. Phys.
  Lett.}\ }\textbf {\bibinfo {volume} {264}},\ \bibinfo {pages} {573} (\bibinfo
  {year} {1997})}\BibitemShut {NoStop}%
\bibitem [{\citenamefont {Chelikowsky}, \citenamefont {Kronik},\ and\
  \citenamefont {Vasiliev}(2003)}]{Chelikowsky2003}%
  \BibitemOpen
  \bibfield  {author} {\bibinfo {author} {\bibfnamefont {J.~R.}\ \bibnamefont
  {Chelikowsky}}, \bibinfo {author} {\bibfnamefont {L.}~\bibnamefont {Kronik}},
  \ and\ \bibinfo {author} {\bibfnamefont {I.}~\bibnamefont {Vasiliev}},\
  }\href@noop {} {\bibfield  {journal} {\bibinfo  {journal} {J. Phys. Condes.
  Matrer}\ }\textbf {\bibinfo {volume} {15}},\ \bibinfo {pages} {R1517}
  (\bibinfo {year} {2003})}\BibitemShut {NoStop}%
\bibitem [{\citenamefont {Gavnholt}\ \emph {et~al.}(2009)\citenamefont
  {Gavnholt}, \citenamefont {Rubio}, \citenamefont {Olsen}, \citenamefont
  {Thygesen},\ and\ \citenamefont {Schiotz}}]{Gavnholt2009}%
  \BibitemOpen
  \bibfield  {author} {\bibinfo {author} {\bibfnamefont {J.}~\bibnamefont
  {Gavnholt}}, \bibinfo {author} {\bibfnamefont {A.}~\bibnamefont {Rubio}},
  \bibinfo {author} {\bibfnamefont {T.}~\bibnamefont {Olsen}}, \bibinfo
  {author} {\bibfnamefont {K.~S.}\ \bibnamefont {Thygesen}}, \ and\ \bibinfo
  {author} {\bibfnamefont {J.}~\bibnamefont {Schiotz}},\ }\href@noop {}
  {\bibfield  {journal} {\bibinfo  {journal} {Phys. Rev. B}\ }\textbf {\bibinfo
  {volume} {79}},\ \bibinfo {pages} {195405} (\bibinfo {year}
  {2009})}\BibitemShut {NoStop}%
\bibitem [{\citenamefont {Hirata}\ and\ \citenamefont
  {Head-Gordon}(1999{\natexlab{a}})}]{Hirata1999}%
  \BibitemOpen
  \bibfield  {author} {\bibinfo {author} {\bibfnamefont {S.}~\bibnamefont
  {Hirata}}\ and\ \bibinfo {author} {\bibfnamefont {M.}~\bibnamefont
  {Head-Gordon}},\ }\href@noop {} {\bibfield  {journal} {\bibinfo  {journal}
  {Chem. Phys. Lett.}\ }\textbf {\bibinfo {volume} {302}},\ \bibinfo {pages}
  {375} (\bibinfo {year} {1999}{\natexlab{a}})}\BibitemShut {NoStop}%
\bibitem [{\citenamefont {Hirata}, \citenamefont {Lee},\ and\ \citenamefont
  {Head-Gordon}(1999)}]{Hirata1999c}%
  \BibitemOpen
  \bibfield  {author} {\bibinfo {author} {\bibfnamefont {S.}~\bibnamefont
  {Hirata}}, \bibinfo {author} {\bibfnamefont {T.~J.}\ \bibnamefont {Lee}}, \
  and\ \bibinfo {author} {\bibfnamefont {M.}~\bibnamefont {Head-Gordon}},\
  }\href@noop {} {\bibfield  {journal} {\bibinfo  {journal} {J. Chem. Phys.}\
  }\textbf {\bibinfo {volume} {111}},\ \bibinfo {pages} {8904} (\bibinfo {year}
  {1999})}\BibitemShut {NoStop}%
\bibitem [{\citenamefont {Jacquemin}\ \emph {et~al.}(2008)\citenamefont
  {Jacquemin}, \citenamefont {Perpete}, \citenamefont {Scuseria}, \citenamefont
  {Ciofini},\ and\ \citenamefont {Adamo}}]{Jacquemin2008}%
  \BibitemOpen
  \bibfield  {author} {\bibinfo {author} {\bibfnamefont {D.}~\bibnamefont
  {Jacquemin}}, \bibinfo {author} {\bibfnamefont {E.~A.}\ \bibnamefont
  {Perpete}}, \bibinfo {author} {\bibfnamefont {G.~E.}\ \bibnamefont
  {Scuseria}}, \bibinfo {author} {\bibfnamefont {I.}~\bibnamefont {Ciofini}}, \
  and\ \bibinfo {author} {\bibfnamefont {C.}~\bibnamefont {Adamo}},\
  }\href@noop {} {\bibfield  {journal} {\bibinfo  {journal} {J. Chem. Theory
  Comput.}\ }\textbf {\bibinfo {volume} {4}},\ \bibinfo {pages} {123} (\bibinfo
  {year} {2008})}\BibitemShut {NoStop}%
\bibitem [{\citenamefont {Stein}, \citenamefont {Kronik},\ and\ \citenamefont
  {Baer}(2009{\natexlab{a}})}]{Stein2009}%
  \BibitemOpen
  \bibfield  {author} {\bibinfo {author} {\bibfnamefont {T.}~\bibnamefont
  {Stein}}, \bibinfo {author} {\bibfnamefont {L.}~\bibnamefont {Kronik}}, \
  and\ \bibinfo {author} {\bibfnamefont {R.}~\bibnamefont {Baer}},\ }\href@noop
  {} {\bibfield  {journal} {\bibinfo  {journal} {J. Chem. Phys.}\ }\textbf
  {\bibinfo {volume} {131}},\ \bibinfo {pages} {244119} (\bibinfo {year}
  {2009}{\natexlab{a}})}\BibitemShut {NoStop}%
\bibitem [{\citenamefont {Stein}, \citenamefont {Kronik},\ and\ \citenamefont
  {Baer}(2009{\natexlab{b}})}]{Stein2009a}%
  \BibitemOpen
  \bibfield  {author} {\bibinfo {author} {\bibfnamefont {T.}~\bibnamefont
  {Stein}}, \bibinfo {author} {\bibfnamefont {L.}~\bibnamefont {Kronik}}, \
  and\ \bibinfo {author} {\bibfnamefont {R.}~\bibnamefont {Baer}},\ }\href@noop
  {} {\bibfield  {journal} {\bibinfo  {journal} {J. Am. Chem. Soc.}\ }\textbf
  {\bibinfo {volume} {131}},\ \bibinfo {pages} {2818} (\bibinfo {year}
  {2009}{\natexlab{b}})}\BibitemShut {NoStop}%
\bibitem [{\citenamefont {Phillips}\ \emph {et~al.}(2011)\citenamefont
  {Phillips}, \citenamefont {Zheng}, \citenamefont {Hyla}, \citenamefont
  {Laine}, \citenamefont {Goodson}, \citenamefont {Geva},\ and\ \citenamefont
  {Dunietz}}]{Phillips2011}%
  \BibitemOpen
  \bibfield  {author} {\bibinfo {author} {\bibfnamefont {H.}~\bibnamefont
  {Phillips}}, \bibinfo {author} {\bibfnamefont {S.}~\bibnamefont {Zheng}},
  \bibinfo {author} {\bibfnamefont {A.}~\bibnamefont {Hyla}}, \bibinfo {author}
  {\bibfnamefont {R.}~\bibnamefont {Laine}}, \bibinfo {author} {\bibfnamefont
  {T.}~\bibnamefont {Goodson}}, \bibinfo {author} {\bibfnamefont
  {E.}~\bibnamefont {Geva}}, \ and\ \bibinfo {author} {\bibfnamefont {B.~D.}\
  \bibnamefont {Dunietz}},\ }\href@noop {} {\bibfield  {journal} {\bibinfo
  {journal} {J. Phys. Chem. A}\ }\textbf {\bibinfo {volume} {116}},\ \bibinfo
  {pages} {1137} (\bibinfo {year} {2011})}\BibitemShut {NoStop}%
\bibitem [{\citenamefont {Ou}\ \emph {et~al.}(2014)\citenamefont {Ou},
  \citenamefont {Fatehi}, \citenamefont {Alguire}, \citenamefont {Shao},\ and\
  \citenamefont {Subotnik}}]{Ou2014}%
  \BibitemOpen
  \bibfield  {author} {\bibinfo {author} {\bibfnamefont {Q.}~\bibnamefont
  {Ou}}, \bibinfo {author} {\bibfnamefont {S.}~\bibnamefont {Fatehi}}, \bibinfo
  {author} {\bibfnamefont {E.}~\bibnamefont {Alguire}}, \bibinfo {author}
  {\bibfnamefont {Y.}~\bibnamefont {Shao}}, \ and\ \bibinfo {author}
  {\bibfnamefont {J.~E.}\ \bibnamefont {Subotnik}},\ }\href
  {http://scitation.aip.org/content/aip/journal/jcp/141/2/10.1063/1.4887256}
  {\bibfield  {journal} {\bibinfo  {journal} {J. Chem. Phys.}\ }\textbf
  {\bibinfo {volume} {141}},\ \bibinfo {pages} {024114} (\bibinfo {year}
  {2014})}\BibitemShut {NoStop}%
\bibitem [{\citenamefont {Parac}\ and\ \citenamefont
  {Grimme}(2003)}]{Parac2003}%
  \BibitemOpen
  \bibfield  {author} {\bibinfo {author} {\bibfnamefont {M.}~\bibnamefont
  {Parac}}\ and\ \bibinfo {author} {\bibfnamefont {S.}~\bibnamefont {Grimme}},\
  }\href@noop {} {\bibfield  {journal} {\bibinfo  {journal} {Chem. Phys.}\
  }\textbf {\bibinfo {volume} {292}},\ \bibinfo {pages} {11} (\bibinfo {year}
  {2003})}\BibitemShut {NoStop}%
\bibitem [{\citenamefont {Grimme}\ and\ \citenamefont
  {Parac}(2003)}]{Grimme2003}%
  \BibitemOpen
  \bibfield  {author} {\bibinfo {author} {\bibfnamefont {S.}~\bibnamefont
  {Grimme}}\ and\ \bibinfo {author} {\bibfnamefont {M.}~\bibnamefont {Parac}},\
  }\href@noop {} {\bibfield  {journal} {\bibinfo  {journal} {Chem. Phys.
  Chem.}\ }\textbf {\bibinfo {volume} {4}},\ \bibinfo {pages} {292} (\bibinfo
  {year} {2003})}\BibitemShut {NoStop}%
\bibitem [{\citenamefont {Dreuw}\ and\ \citenamefont
  {Head-Gordon}(2004)}]{Dreuw2004}%
  \BibitemOpen
  \bibfield  {author} {\bibinfo {author} {\bibfnamefont {A.}~\bibnamefont
  {Dreuw}}\ and\ \bibinfo {author} {\bibfnamefont {M.}~\bibnamefont
  {Head-Gordon}},\ }\href@noop {} {\bibfield  {journal} {\bibinfo  {journal}
  {J. Am. Chem. Soc.}\ }\textbf {\bibinfo {volume} {126}},\ \bibinfo {pages}
  {4007} (\bibinfo {year} {2004})}\BibitemShut {NoStop}%
\bibitem [{\citenamefont {Maitra}\ \emph {et~al.}(2004)\citenamefont {Maitra},
  \citenamefont {Zhang}, \citenamefont {Cave},\ and\ \citenamefont
  {Burke}}]{Maitra2004}%
  \BibitemOpen
  \bibfield  {author} {\bibinfo {author} {\bibfnamefont {N.~T.}\ \bibnamefont
  {Maitra}}, \bibinfo {author} {\bibfnamefont {F.}~\bibnamefont {Zhang}},
  \bibinfo {author} {\bibfnamefont {R.~J.}\ \bibnamefont {Cave}}, \ and\
  \bibinfo {author} {\bibfnamefont {K.}~\bibnamefont {Burke}},\ }\href@noop {}
  {\bibfield  {journal} {\bibinfo  {journal} {J. Chem. Phys.}\ }\textbf
  {\bibinfo {volume} {120}},\ \bibinfo {pages} {5932} (\bibinfo {year}
  {2004})}\BibitemShut {NoStop}%
\bibitem [{\citenamefont {Hieringer}\ and\ \citenamefont
  {G{\"o}rling}(2006)}]{Hieringer2006}%
  \BibitemOpen
  \bibfield  {author} {\bibinfo {author} {\bibfnamefont {W.}~\bibnamefont
  {Hieringer}}\ and\ \bibinfo {author} {\bibfnamefont {A.}~\bibnamefont
  {G{\"o}rling}},\ }\href@noop {} {\bibfield  {journal} {\bibinfo  {journal}
  {Chem. Phys. Lett.}\ }\textbf {\bibinfo {volume} {419}},\ \bibinfo {pages}
  {557} (\bibinfo {year} {2006})}\BibitemShut {NoStop}%
\bibitem [{\citenamefont {Levine}\ \emph {et~al.}(2006)\citenamefont {Levine},
  \citenamefont {Ko}, \citenamefont {Quenneville},\ and\ \citenamefont
  {Martinez}}]{Levine2006}%
  \BibitemOpen
  \bibfield  {author} {\bibinfo {author} {\bibfnamefont {B.~G.}\ \bibnamefont
  {Levine}}, \bibinfo {author} {\bibfnamefont {C.}~\bibnamefont {Ko}}, \bibinfo
  {author} {\bibfnamefont {J.}~\bibnamefont {Quenneville}}, \ and\ \bibinfo
  {author} {\bibfnamefont {T.~J.}\ \bibnamefont {Martinez}},\ }\href@noop {}
  {\bibfield  {journal} {\bibinfo  {journal} {Mol. Phys.}\ }\textbf {\bibinfo
  {volume} {104}},\ \bibinfo {pages} {1039} (\bibinfo {year}
  {2006})}\BibitemShut {NoStop}%
\bibitem [{\citenamefont {Lopata}\ \emph {et~al.}(2011)\citenamefont {Lopata},
  \citenamefont {Reslan}, \citenamefont {Kowaska}, \citenamefont {Neuhauser},
  \citenamefont {Govind},\ and\ \citenamefont {Kowalski}}]{Lopata2011}%
  \BibitemOpen
  \bibfield  {author} {\bibinfo {author} {\bibfnamefont {K.}~\bibnamefont
  {Lopata}}, \bibinfo {author} {\bibfnamefont {R.}~\bibnamefont {Reslan}},
  \bibinfo {author} {\bibfnamefont {M.}~\bibnamefont {Kowaska}}, \bibinfo
  {author} {\bibfnamefont {D.}~\bibnamefont {Neuhauser}}, \bibinfo {author}
  {\bibfnamefont {N.}~\bibnamefont {Govind}}, \ and\ \bibinfo {author}
  {\bibfnamefont {K.}~\bibnamefont {Kowalski}},\ }\href@noop {} {\bibfield
  {journal} {\bibinfo  {journal} {J. Chem. Theory Comput.}\ }\textbf {\bibinfo
  {volume} {7}},\ \bibinfo {pages} {3686} (\bibinfo {year} {2011})}\BibitemShut
  {NoStop}%
\bibitem [{\citenamefont {Kowalczyk}, \citenamefont {Yost},\ and\ \citenamefont
  {Van~Voorhis}(2011)}]{Kowalczyk2011}%
  \BibitemOpen
  \bibfield  {author} {\bibinfo {author} {\bibfnamefont {T.}~\bibnamefont
  {Kowalczyk}}, \bibinfo {author} {\bibfnamefont {S.~R.}\ \bibnamefont {Yost}},
  \ and\ \bibinfo {author} {\bibfnamefont {T.}~\bibnamefont {Van~Voorhis}},\
  }\href
  {http://scitation.aip.org/content/aip/journal/jcp/134/5/10.1063/1.3530801}
  {\bibfield  {journal} {\bibinfo  {journal} {J. Chem. Phys.}\ }\textbf
  {\bibinfo {volume} {134}},\ \bibinfo {pages} {054128} (\bibinfo {year}
  {2011})}\BibitemShut {NoStop}%
\bibitem [{\citenamefont {Isborn}\ \emph {et~al.}(2013)\citenamefont {Isborn},
  \citenamefont {Mar}, \citenamefont {Curchod}, \citenamefont {Tavernelli},\
  and\ \citenamefont {Mart{\'\i}nez}}]{Isborn2013}%
  \BibitemOpen
  \bibfield  {author} {\bibinfo {author} {\bibfnamefont {C.~M.}\ \bibnamefont
  {Isborn}}, \bibinfo {author} {\bibfnamefont {B.~D.}\ \bibnamefont {Mar}},
  \bibinfo {author} {\bibfnamefont {B.~F.}\ \bibnamefont {Curchod}}, \bibinfo
  {author} {\bibfnamefont {I.}~\bibnamefont {Tavernelli}}, \ and\ \bibinfo
  {author} {\bibfnamefont {T.~J.}\ \bibnamefont {Mart{\'\i}nez}},\ }\href
  {http://pubs.acs.org/doi/abs/10.1021/jp4058274} {\bibfield  {journal}
  {\bibinfo  {journal} {J. Phys. Chem. B}\ }\textbf {\bibinfo {volume} {117}},\
  \bibinfo {pages} {12189} (\bibinfo {year} {2013})}\BibitemShut {NoStop}%
\bibitem [{\citenamefont {Albrecht}\ \emph {et~al.}(1998)\citenamefont
  {Albrecht}, \citenamefont {Reining}, \citenamefont {Del~Sole},\ and\
  \citenamefont {Onida}}]{Albrecht1998}%
  \BibitemOpen
  \bibfield  {author} {\bibinfo {author} {\bibfnamefont {S.}~\bibnamefont
  {Albrecht}}, \bibinfo {author} {\bibfnamefont {L.}~\bibnamefont {Reining}},
  \bibinfo {author} {\bibfnamefont {R.}~\bibnamefont {Del~Sole}}, \ and\
  \bibinfo {author} {\bibfnamefont {G.}~\bibnamefont {Onida}},\ }\href@noop {}
  {\bibfield  {journal} {\bibinfo  {journal} {Phys. Rev. Lett.}\ }\textbf
  {\bibinfo {volume} {80}},\ \bibinfo {pages} {4510} (\bibinfo {year}
  {1998})}\BibitemShut {NoStop}%
\bibitem [{\citenamefont {Rohlfing}\ and\ \citenamefont
  {Louie}(2000)}]{Rohlfing2000}%
  \BibitemOpen
  \bibfield  {author} {\bibinfo {author} {\bibfnamefont {M.}~\bibnamefont
  {Rohlfing}}\ and\ \bibinfo {author} {\bibfnamefont {S.~G.}\ \bibnamefont
  {Louie}},\ }\href@noop {} {\bibfield  {journal} {\bibinfo  {journal} {Phys.
  Rev. B}\ }\textbf {\bibinfo {volume} {62}},\ \bibinfo {pages} {4927}
  (\bibinfo {year} {2000})}\BibitemShut {NoStop}%
\bibitem [{\citenamefont {Sottile}\ \emph {et~al.}(2007)\citenamefont
  {Sottile}, \citenamefont {Marsili}, \citenamefont {Olevano},\ and\
  \citenamefont {Reining}}]{Sottile2007}%
  \BibitemOpen
  \bibfield  {author} {\bibinfo {author} {\bibfnamefont {F.}~\bibnamefont
  {Sottile}}, \bibinfo {author} {\bibfnamefont {M.}~\bibnamefont {Marsili}},
  \bibinfo {author} {\bibfnamefont {V.}~\bibnamefont {Olevano}}, \ and\
  \bibinfo {author} {\bibfnamefont {L.}~\bibnamefont {Reining}},\ }\href@noop
  {} {\bibfield  {journal} {\bibinfo  {journal} {Phys. Rev. B}\ }\textbf
  {\bibinfo {volume} {76}},\ \bibinfo {pages} {161103} (\bibinfo {year}
  {2007})}\BibitemShut {NoStop}%
\bibitem [{\citenamefont {Ramos}\ \emph {et~al.}(2008)\citenamefont {Ramos},
  \citenamefont {Paier}, \citenamefont {Kresse},\ and\ \citenamefont
  {Bechstedt}}]{Ramos2008}%
  \BibitemOpen
  \bibfield  {author} {\bibinfo {author} {\bibfnamefont {L.}~\bibnamefont
  {Ramos}}, \bibinfo {author} {\bibfnamefont {J.}~\bibnamefont {Paier}},
  \bibinfo {author} {\bibfnamefont {G.}~\bibnamefont {Kresse}}, \ and\ \bibinfo
  {author} {\bibfnamefont {F.}~\bibnamefont {Bechstedt}},\ }\href@noop {}
  {\bibfield  {journal} {\bibinfo  {journal} {Phys. Rev. B}\ }\textbf {\bibinfo
  {volume} {78}},\ \bibinfo {pages} {195423} (\bibinfo {year}
  {2008})}\BibitemShut {NoStop}%
\bibitem [{\citenamefont {Rocca}\ \emph {et~al.}(2012)\citenamefont {Rocca},
  \citenamefont {Ping}, \citenamefont {Gebauer},\ and\ \citenamefont
  {Galli}}]{Rocca2012a}%
  \BibitemOpen
  \bibfield  {author} {\bibinfo {author} {\bibfnamefont {D.}~\bibnamefont
  {Rocca}}, \bibinfo {author} {\bibfnamefont {Y.}~\bibnamefont {Ping}},
  \bibinfo {author} {\bibfnamefont {R.}~\bibnamefont {Gebauer}}, \ and\
  \bibinfo {author} {\bibfnamefont {G.}~\bibnamefont {Galli}},\ }\href@noop {}
  {\bibfield  {journal} {\bibinfo  {journal} {Phys. Rev. B}\ }\textbf {\bibinfo
  {volume} {85}},\ \bibinfo {pages} {045116} (\bibinfo {year}
  {2012})}\BibitemShut {NoStop}%
\bibitem [{\citenamefont {Hedin}(1965)}]{Hedin1965}%
  \BibitemOpen
  \bibfield  {author} {\bibinfo {author} {\bibfnamefont {L.}~\bibnamefont
  {Hedin}},\ }\href@noop {} {\bibfield  {journal} {\bibinfo  {journal} {Phys.
  Rev.}\ }\textbf {\bibinfo {volume} {139}},\ \bibinfo {pages} {A796} (\bibinfo
  {year} {1965})}\BibitemShut {NoStop}%
\bibitem [{\citenamefont {Salpeter}\ and\ \citenamefont
  {Bethe}(1951)}]{Salpeter1951}%
  \BibitemOpen
  \bibfield  {author} {\bibinfo {author} {\bibfnamefont {E.~E.}\ \bibnamefont
  {Salpeter}}\ and\ \bibinfo {author} {\bibfnamefont {H.~A.}\ \bibnamefont
  {Bethe}},\ }\href@noop {} {\bibfield  {journal} {\bibinfo  {journal} {Phys.
  Rev.}\ }\textbf {\bibinfo {volume} {84}},\ \bibinfo {pages} {1232} (\bibinfo
  {year} {1951})}\BibitemShut {NoStop}%
\bibitem [{\citenamefont {Hybertsen}\ and\ \citenamefont
  {Louie}(1985)}]{Hybertsen1985}%
  \BibitemOpen
  \bibfield  {author} {\bibinfo {author} {\bibfnamefont {M.~S.}\ \bibnamefont
  {Hybertsen}}\ and\ \bibinfo {author} {\bibfnamefont {S.~G.}\ \bibnamefont
  {Louie}},\ }\href@noop {} {\bibfield  {journal} {\bibinfo  {journal} {Phys.
  Rev. Lett.}\ }\textbf {\bibinfo {volume} {55}},\ \bibinfo {pages} {1418}
  (\bibinfo {year} {1985})}\BibitemShut {NoStop}%
\bibitem [{\citenamefont {Hybertsen}\ and\ \citenamefont
  {Louie}(1986)}]{Hybertsen1986}%
  \BibitemOpen
  \bibfield  {author} {\bibinfo {author} {\bibfnamefont {M.~S.}\ \bibnamefont
  {Hybertsen}}\ and\ \bibinfo {author} {\bibfnamefont {S.~G.}\ \bibnamefont
  {Louie}},\ }\href@noop {} {\bibfield  {journal} {\bibinfo  {journal} {Phys.
  Rev. B}\ }\textbf {\bibinfo {volume} {34}},\ \bibinfo {pages} {5390}
  (\bibinfo {year} {1986})}\BibitemShut {NoStop}%
\bibitem [{\citenamefont {Steinbeck}\ \emph {et~al.}(1999)\citenamefont
  {Steinbeck}, \citenamefont {Rubio}, \citenamefont {Reining}, \citenamefont
  {Torrent}, \citenamefont {White},\ and\ \citenamefont
  {Godby}}]{Steinbeck1999}%
  \BibitemOpen
  \bibfield  {author} {\bibinfo {author} {\bibfnamefont {L.}~\bibnamefont
  {Steinbeck}}, \bibinfo {author} {\bibfnamefont {A.}~\bibnamefont {Rubio}},
  \bibinfo {author} {\bibfnamefont {L.}~\bibnamefont {Reining}}, \bibinfo
  {author} {\bibfnamefont {M.}~\bibnamefont {Torrent}}, \bibinfo {author}
  {\bibfnamefont {I.}~\bibnamefont {White}}, \ and\ \bibinfo {author}
  {\bibfnamefont {R.}~\bibnamefont {Godby}},\ }\href@noop {} {\bibfield
  {journal} {\bibinfo  {journal} {Comput. Phys. Commun.}\ }\textbf {\bibinfo
  {volume} {125}},\ \bibinfo {pages} {05} (\bibinfo {year} {1999})}\BibitemShut
  {NoStop}%
\bibitem [{\citenamefont {Rieger}\ \emph {et~al.}(1999)\citenamefont {Rieger},
  \citenamefont {Steinbeck}, \citenamefont {White}, \citenamefont {Rojas},\
  and\ \citenamefont {Godby}}]{Rieger1999}%
  \BibitemOpen
  \bibfield  {author} {\bibinfo {author} {\bibfnamefont {M.~M.}\ \bibnamefont
  {Rieger}}, \bibinfo {author} {\bibfnamefont {L.}~\bibnamefont {Steinbeck}},
  \bibinfo {author} {\bibfnamefont {I.}~\bibnamefont {White}}, \bibinfo
  {author} {\bibfnamefont {H.}~\bibnamefont {Rojas}}, \ and\ \bibinfo {author}
  {\bibfnamefont {R.}~\bibnamefont {Godby}},\ }\href@noop {} {\bibfield
  {journal} {\bibinfo  {journal} {Comput. Phys. Commun.}\ }\textbf {\bibinfo
  {volume} {117}},\ \bibinfo {pages} {211} (\bibinfo {year}
  {1999})}\BibitemShut {NoStop}%
\bibitem [{\citenamefont {Rinke}\ \emph {et~al.}(2005)\citenamefont {Rinke},
  \citenamefont {Qteish}, \citenamefont {Neugebauer}, \citenamefont
  {Freysoldt},\ and\ \citenamefont {Scheffler}}]{Rinke2005}%
  \BibitemOpen
  \bibfield  {author} {\bibinfo {author} {\bibfnamefont {P.}~\bibnamefont
  {Rinke}}, \bibinfo {author} {\bibfnamefont {A.}~\bibnamefont {Qteish}},
  \bibinfo {author} {\bibfnamefont {J.}~\bibnamefont {Neugebauer}}, \bibinfo
  {author} {\bibfnamefont {C.}~\bibnamefont {Freysoldt}}, \ and\ \bibinfo
  {author} {\bibfnamefont {M.}~\bibnamefont {Scheffler}},\ }\href@noop {}
  {\bibfield  {journal} {\bibinfo  {journal} {New J. Phys.}\ }\textbf {\bibinfo
  {volume} {7}},\  (\bibinfo {year} {2005})}\BibitemShut {NoStop}%
\bibitem [{\citenamefont {Neaton}, \citenamefont {Hybertsen},\ and\
  \citenamefont {Louie}(2006)}]{Neaton2006}%
  \BibitemOpen
  \bibfield  {author} {\bibinfo {author} {\bibfnamefont {J.~B.}\ \bibnamefont
  {Neaton}}, \bibinfo {author} {\bibfnamefont {M.~S.}\ \bibnamefont
  {Hybertsen}}, \ and\ \bibinfo {author} {\bibfnamefont {S.~G.}\ \bibnamefont
  {Louie}},\ }\href@noop {} {\bibfield  {journal} {\bibinfo  {journal} {Phys.
  Rev. Lett.}\ }\textbf {\bibinfo {volume} {97}},\ \bibinfo {pages} {216405}
  (\bibinfo {year} {2006})}\BibitemShut {NoStop}%
\bibitem [{\citenamefont {Tiago}\ and\ \citenamefont
  {Chelikowsky}(2006)}]{Tiago2006}%
  \BibitemOpen
  \bibfield  {author} {\bibinfo {author} {\bibfnamefont {M.~L.}\ \bibnamefont
  {Tiago}}\ and\ \bibinfo {author} {\bibfnamefont {J.~R.}\ \bibnamefont
  {Chelikowsky}},\ }\href@noop {} {\bibfield  {journal} {\bibinfo  {journal}
  {Phys. Rev. B}\ }\textbf {\bibinfo {volume} {73}},\ \bibinfo {pages} {205334}
  (\bibinfo {year} {2006})}\BibitemShut {NoStop}%
\bibitem [{\citenamefont {Friedrich}\ and\ \citenamefont
  {Schindlmayr}(2006)}]{Friedrich2006}%
  \BibitemOpen
  \bibfield  {author} {\bibinfo {author} {\bibfnamefont {C.}~\bibnamefont
  {Friedrich}}\ and\ \bibinfo {author} {\bibfnamefont {A.}~\bibnamefont
  {Schindlmayr}},\ }\href@noop {} {\bibfield  {journal} {\bibinfo  {journal}
  {NIC Series}\ }\textbf {\bibinfo {volume} {31}},\ \bibinfo {pages} {335}
  (\bibinfo {year} {2006})}\BibitemShut {NoStop}%
\bibitem [{\citenamefont {Gruning}, \citenamefont {Marini},\ and\ \citenamefont
  {Rubio}(2006)}]{Gruning2006}%
  \BibitemOpen
  \bibfield  {author} {\bibinfo {author} {\bibfnamefont {M.}~\bibnamefont
  {Gruning}}, \bibinfo {author} {\bibfnamefont {A.}~\bibnamefont {Marini}}, \
  and\ \bibinfo {author} {\bibfnamefont {A.}~\bibnamefont {Rubio}},\
  }\href@noop {} {\bibfield  {journal} {\bibinfo  {journal} {J. Chem. Phys.}\
  }\textbf {\bibinfo {volume} {124}},\ \bibinfo {pages} {154108} (\bibinfo
  {year} {2006})}\BibitemShut {NoStop}%
\bibitem [{\citenamefont {Shishkin}\ and\ \citenamefont
  {Kresse}(2007)}]{Shishkin2007}%
  \BibitemOpen
  \bibfield  {author} {\bibinfo {author} {\bibfnamefont {M.}~\bibnamefont
  {Shishkin}}\ and\ \bibinfo {author} {\bibfnamefont {G.}~\bibnamefont
  {Kresse}},\ }\href@noop {} {\bibfield  {journal} {\bibinfo  {journal} {Phys.
  Rev. B}\ }\textbf {\bibinfo {volume} {75}},\ \bibinfo {pages} {235102}
  (\bibinfo {year} {2007})}\BibitemShut {NoStop}%
\bibitem [{\citenamefont {Huang}\ and\ \citenamefont
  {Carter}(2008)}]{Huang2008}%
  \BibitemOpen
  \bibfield  {author} {\bibinfo {author} {\bibfnamefont {P.}~\bibnamefont
  {Huang}}\ and\ \bibinfo {author} {\bibfnamefont {E.~A.}\ \bibnamefont
  {Carter}},\ }\href@noop {} {\bibfield  {journal} {\bibinfo  {journal} {Annu.
  Rev. Phys. Chem.}\ }\textbf {\bibinfo {volume} {59}},\ \bibinfo {pages} {261}
  (\bibinfo {year} {2008})}\BibitemShut {NoStop}%
\bibitem [{\citenamefont {Rostgaard}, \citenamefont {Jacobsen},\ and\
  \citenamefont {Thygesen}(2010)}]{Rostgaard2010}%
  \BibitemOpen
  \bibfield  {author} {\bibinfo {author} {\bibfnamefont {C.}~\bibnamefont
  {Rostgaard}}, \bibinfo {author} {\bibfnamefont {K.~W.}\ \bibnamefont
  {Jacobsen}}, \ and\ \bibinfo {author} {\bibfnamefont {K.~S.}\ \bibnamefont
  {Thygesen}},\ }\href@noop {} {\bibfield  {journal} {\bibinfo  {journal}
  {Phys. Rev. B}\ }\textbf {\bibinfo {volume} {81}},\ \bibinfo {pages} {085103}
  (\bibinfo {year} {2010})}\BibitemShut {NoStop}%
\bibitem [{\citenamefont {Tamblyn}\ \emph {et~al.}(2011)\citenamefont
  {Tamblyn}, \citenamefont {Darancet}, \citenamefont {Quek}, \citenamefont
  {Bonev},\ and\ \citenamefont {Neaton}}]{Tamblyn2011}%
  \BibitemOpen
  \bibfield  {author} {\bibinfo {author} {\bibfnamefont {I.}~\bibnamefont
  {Tamblyn}}, \bibinfo {author} {\bibfnamefont {P.}~\bibnamefont {Darancet}},
  \bibinfo {author} {\bibfnamefont {S.~Y.}\ \bibnamefont {Quek}}, \bibinfo
  {author} {\bibfnamefont {S.~A.}\ \bibnamefont {Bonev}}, \ and\ \bibinfo
  {author} {\bibfnamefont {J.~B.}\ \bibnamefont {Neaton}},\ }\href@noop {}
  {\bibfield  {journal} {\bibinfo  {journal} {Phys. Rev. B}\ }\textbf {\bibinfo
  {volume} {84}},\ \bibinfo {pages} {201402} (\bibinfo {year}
  {2011})}\BibitemShut {NoStop}%
\bibitem [{\citenamefont {Liao}\ and\ \citenamefont {Carter}(2011)}]{Liao2011}%
  \BibitemOpen
  \bibfield  {author} {\bibinfo {author} {\bibfnamefont {P.}~\bibnamefont
  {Liao}}\ and\ \bibinfo {author} {\bibfnamefont {E.~A.}\ \bibnamefont
  {Carter}},\ }\href@noop {} {\bibfield  {journal} {\bibinfo  {journal} {Phys.
  Chem. Chem. Phys.}\ }\textbf {\bibinfo {volume} {13}},\ \bibinfo {pages}
  {15189} (\bibinfo {year} {2011})}\BibitemShut {NoStop}%
\bibitem [{\citenamefont {Refaely-Abramson}, \citenamefont {Baer},\ and\
  \citenamefont {Kronik}(2011)}]{Refaely-Abramson2011}%
  \BibitemOpen
  \bibfield  {author} {\bibinfo {author} {\bibfnamefont {S.}~\bibnamefont
  {Refaely-Abramson}}, \bibinfo {author} {\bibfnamefont {R.}~\bibnamefont
  {Baer}}, \ and\ \bibinfo {author} {\bibfnamefont {L.}~\bibnamefont
  {Kronik}},\ }\href@noop {} {\bibfield  {journal} {\bibinfo  {journal} {Phys.
  Rev. B}\ }\textbf {\bibinfo {volume} {84}},\ \bibinfo {pages} {075144}
  (\bibinfo {year} {2011})}\BibitemShut {NoStop}%
\bibitem [{\citenamefont {Marom}\ \emph {et~al.}(2012)\citenamefont {Marom},
  \citenamefont {Caruso}, \citenamefont {Ren}, \citenamefont {Hofmann},
  \citenamefont {K{\"o}rzd{\"o}rfer}, \citenamefont {Chelikowsky},
  \citenamefont {Rubio}, \citenamefont {Scheffler},\ and\ \citenamefont
  {Rinke}}]{Marom2012}%
  \BibitemOpen
  \bibfield  {author} {\bibinfo {author} {\bibfnamefont {N.}~\bibnamefont
  {Marom}}, \bibinfo {author} {\bibfnamefont {F.}~\bibnamefont {Caruso}},
  \bibinfo {author} {\bibfnamefont {X.}~\bibnamefont {Ren}}, \bibinfo {author}
  {\bibfnamefont {O.~T.}\ \bibnamefont {Hofmann}}, \bibinfo {author}
  {\bibfnamefont {T.}~\bibnamefont {K{\"o}rzd{\"o}rfer}}, \bibinfo {author}
  {\bibfnamefont {J.~R.}\ \bibnamefont {Chelikowsky}}, \bibinfo {author}
  {\bibfnamefont {A.}~\bibnamefont {Rubio}}, \bibinfo {author} {\bibfnamefont
  {M.}~\bibnamefont {Scheffler}}, \ and\ \bibinfo {author} {\bibfnamefont
  {P.}~\bibnamefont {Rinke}},\ }\href@noop {} {\bibfield  {journal} {\bibinfo
  {journal} {Phys. Rev. B}\ }\textbf {\bibinfo {volume} {86}},\ \bibinfo
  {pages} {245127} (\bibinfo {year} {2012})}\BibitemShut {NoStop}%
\bibitem [{\citenamefont {Isseroff}\ and\ \citenamefont
  {Carter}(2012)}]{Isseroff2012}%
  \BibitemOpen
  \bibfield  {author} {\bibinfo {author} {\bibfnamefont {L.~Y.}\ \bibnamefont
  {Isseroff}}\ and\ \bibinfo {author} {\bibfnamefont {E.~A.}\ \bibnamefont
  {Carter}},\ }\href@noop {} {\bibfield  {journal} {\bibinfo  {journal} {Phys.
  Rev. B}\ }\textbf {\bibinfo {volume} {85}},\ \bibinfo {pages} {235142}
  (\bibinfo {year} {2012})}\BibitemShut {NoStop}%
\bibitem [{\citenamefont {Refaely-Abramson}\ \emph {et~al.}(2012)\citenamefont
  {Refaely-Abramson}, \citenamefont {Sharifzadeh}, \citenamefont {Govind},
  \citenamefont {Autschbach}, \citenamefont {Neaton}, \citenamefont {Baer},\
  and\ \citenamefont {Kronik}}]{Refaely-Abramson2012}%
  \BibitemOpen
  \bibfield  {author} {\bibinfo {author} {\bibfnamefont {S.}~\bibnamefont
  {Refaely-Abramson}}, \bibinfo {author} {\bibfnamefont {S.}~\bibnamefont
  {Sharifzadeh}}, \bibinfo {author} {\bibfnamefont {N.}~\bibnamefont {Govind}},
  \bibinfo {author} {\bibfnamefont {J.}~\bibnamefont {Autschbach}}, \bibinfo
  {author} {\bibfnamefont {J.~B.}\ \bibnamefont {Neaton}}, \bibinfo {author}
  {\bibfnamefont {R.}~\bibnamefont {Baer}}, \ and\ \bibinfo {author}
  {\bibfnamefont {L.}~\bibnamefont {Kronik}},\ }\href@noop {} {\bibfield
  {journal} {\bibinfo  {journal} {Phys. Rev. Lett.}\ }\textbf {\bibinfo
  {volume} {109}},\ \bibinfo {pages} {226405} (\bibinfo {year}
  {2012})}\BibitemShut {NoStop}%
\bibitem [{\citenamefont {Kronik}\ \emph {et~al.}(2012)\citenamefont {Kronik},
  \citenamefont {Stein}, \citenamefont {Refaely-Abramson},\ and\ \citenamefont
  {Baer}}]{Kronik2012}%
  \BibitemOpen
  \bibfield  {author} {\bibinfo {author} {\bibfnamefont {L.}~\bibnamefont
  {Kronik}}, \bibinfo {author} {\bibfnamefont {T.}~\bibnamefont {Stein}},
  \bibinfo {author} {\bibfnamefont {S.}~\bibnamefont {Refaely-Abramson}}, \
  and\ \bibinfo {author} {\bibfnamefont {R.}~\bibnamefont {Baer}},\ }\href@noop
  {} {\bibfield  {journal} {\bibinfo  {journal} {J. Chem. Theory Comput.}\
  }\textbf {\bibinfo {volume} {8}},\ \bibinfo {pages} {1515} (\bibinfo {year}
  {2012})}\BibitemShut {NoStop}%
\bibitem [{\citenamefont {Benedict}, \citenamefont {Shirley},\ and\
  \citenamefont {Bohn}(1998)}]{Benedict1998}%
  \BibitemOpen
  \bibfield  {author} {\bibinfo {author} {\bibfnamefont {L.~X.}\ \bibnamefont
  {Benedict}}, \bibinfo {author} {\bibfnamefont {E.~L.}\ \bibnamefont
  {Shirley}}, \ and\ \bibinfo {author} {\bibfnamefont {R.~B.}\ \bibnamefont
  {Bohn}},\ }\href@noop {} {\bibfield  {journal} {\bibinfo  {journal} {Phys.
  Rev. Lett.}\ }\textbf {\bibinfo {volume} {80}},\ \bibinfo {pages} {4514}
  (\bibinfo {year} {1998})}\BibitemShut {NoStop}%
\bibitem [{\citenamefont {Benedict}\ \emph {et~al.}(2003)\citenamefont
  {Benedict}, \citenamefont {Puzder}, \citenamefont {Williamson}, \citenamefont
  {Grossman}, \citenamefont {Galli}, \citenamefont {Klepeis}, \citenamefont
  {Raty},\ and\ \citenamefont {Pankratov}}]{Benedict2003}%
  \BibitemOpen
  \bibfield  {author} {\bibinfo {author} {\bibfnamefont {L.~X.}\ \bibnamefont
  {Benedict}}, \bibinfo {author} {\bibfnamefont {A.}~\bibnamefont {Puzder}},
  \bibinfo {author} {\bibfnamefont {A.~J.}\ \bibnamefont {Williamson}},
  \bibinfo {author} {\bibfnamefont {J.~C.}\ \bibnamefont {Grossman}}, \bibinfo
  {author} {\bibfnamefont {G.}~\bibnamefont {Galli}}, \bibinfo {author}
  {\bibfnamefont {J.~E.}\ \bibnamefont {Klepeis}}, \bibinfo {author}
  {\bibfnamefont {J.-Y.}\ \bibnamefont {Raty}}, \ and\ \bibinfo {author}
  {\bibfnamefont {O.}~\bibnamefont {Pankratov}},\ }\href@noop {} {\bibfield
  {journal} {\bibinfo  {journal} {Phys. Rev. B}\ }\textbf {\bibinfo {volume}
  {68}},\ \bibinfo {pages} {085310} (\bibinfo {year} {2003})}\BibitemShut
  {NoStop}%
\bibitem [{\citenamefont {Spataru}\ \emph {et~al.}(2004)\citenamefont
  {Spataru}, \citenamefont {Ismail-Beigi}, \citenamefont {Benedict},\ and\
  \citenamefont {Louie}}]{Spataru2004}%
  \BibitemOpen
  \bibfield  {author} {\bibinfo {author} {\bibfnamefont {C.~D.}\ \bibnamefont
  {Spataru}}, \bibinfo {author} {\bibfnamefont {S.}~\bibnamefont
  {Ismail-Beigi}}, \bibinfo {author} {\bibfnamefont {L.~X.}\ \bibnamefont
  {Benedict}}, \ and\ \bibinfo {author} {\bibfnamefont {S.~G.}\ \bibnamefont
  {Louie}},\ }\href@noop {} {\bibfield  {journal} {\bibinfo  {journal} {Phys.
  Rev. Lett.}\ }\textbf {\bibinfo {volume} {92}},\ \bibinfo {pages} {077402}
  (\bibinfo {year} {2004})}\BibitemShut {NoStop}%
\bibitem [{\citenamefont {Sai}\ \emph {et~al.}(2008)\citenamefont {Sai},
  \citenamefont {Tiago}, \citenamefont {Chelikowsky},\ and\ \citenamefont
  {Reboredo}}]{Sai2008}%
  \BibitemOpen
  \bibfield  {author} {\bibinfo {author} {\bibfnamefont {N.}~\bibnamefont
  {Sai}}, \bibinfo {author} {\bibfnamefont {M.~L.}\ \bibnamefont {Tiago}},
  \bibinfo {author} {\bibfnamefont {J.~R.}\ \bibnamefont {Chelikowsky}}, \ and\
  \bibinfo {author} {\bibfnamefont {F.~A.}\ \bibnamefont {Reboredo}},\
  }\href@noop {} {\bibfield  {journal} {\bibinfo  {journal} {Phys. Rev. B}\
  }\textbf {\bibinfo {volume} {77}},\ \bibinfo {pages} {161306} (\bibinfo
  {year} {2008})}\BibitemShut {NoStop}%
\bibitem [{\citenamefont {Fuchs}\ \emph {et~al.}(2008)\citenamefont {Fuchs},
  \citenamefont {R{\"o}dl}, \citenamefont {Schleife},\ and\ \citenamefont
  {Bechstedt}}]{Fuchs2008}%
  \BibitemOpen
  \bibfield  {author} {\bibinfo {author} {\bibfnamefont {F.}~\bibnamefont
  {Fuchs}}, \bibinfo {author} {\bibfnamefont {C.}~\bibnamefont {R{\"o}dl}},
  \bibinfo {author} {\bibfnamefont {A.}~\bibnamefont {Schleife}}, \ and\
  \bibinfo {author} {\bibfnamefont {F.}~\bibnamefont {Bechstedt}},\ }\href@noop
  {} {\bibfield  {journal} {\bibinfo  {journal} {Phys. Rev. B}\ }\textbf
  {\bibinfo {volume} {78}},\ \bibinfo {pages} {085103} (\bibinfo {year}
  {2008})}\BibitemShut {NoStop}%
\bibitem [{\citenamefont {Palummo}\ \emph {et~al.}(2009)\citenamefont
  {Palummo}, \citenamefont {Hogan}, \citenamefont {Sottile}, \citenamefont
  {Bagala},\ and\ \citenamefont {Rubio}}]{Palummo2009}%
  \BibitemOpen
  \bibfield  {author} {\bibinfo {author} {\bibfnamefont {M.}~\bibnamefont
  {Palummo}}, \bibinfo {author} {\bibfnamefont {C.}~\bibnamefont {Hogan}},
  \bibinfo {author} {\bibfnamefont {F.}~\bibnamefont {Sottile}}, \bibinfo
  {author} {\bibfnamefont {P.}~\bibnamefont {Bagala}}, \ and\ \bibinfo {author}
  {\bibfnamefont {A.}~\bibnamefont {Rubio}},\ }\href@noop {} {\bibfield
  {journal} {\bibinfo  {journal} {J. Chem. Phys.}\ }\textbf {\bibinfo {volume}
  {131}},\ \bibinfo {pages} {084102} (\bibinfo {year} {2009})}\BibitemShut
  {NoStop}%
\bibitem [{\citenamefont {Schimka}\ \emph {et~al.}(2010)\citenamefont
  {Schimka}, \citenamefont {Harl}, \citenamefont {Stroppa}, \citenamefont
  {Gruneis}, \citenamefont {Marsman}, \citenamefont {Mittendorfer},\ and\
  \citenamefont {Kresse}}]{Schimka2010}%
  \BibitemOpen
  \bibfield  {author} {\bibinfo {author} {\bibfnamefont {L.}~\bibnamefont
  {Schimka}}, \bibinfo {author} {\bibfnamefont {J.}~\bibnamefont {Harl}},
  \bibinfo {author} {\bibfnamefont {A.}~\bibnamefont {Stroppa}}, \bibinfo
  {author} {\bibfnamefont {A.}~\bibnamefont {Gruneis}}, \bibinfo {author}
  {\bibfnamefont {M.}~\bibnamefont {Marsman}}, \bibinfo {author} {\bibfnamefont
  {F.}~\bibnamefont {Mittendorfer}}, \ and\ \bibinfo {author} {\bibfnamefont
  {G.}~\bibnamefont {Kresse}},\ }\href@noop {} {\bibfield  {journal} {\bibinfo
  {journal} {Nat. Mater.}\ }\textbf {\bibinfo {volume} {9}},\ \bibinfo {pages}
  {741} (\bibinfo {year} {2010})}\BibitemShut {NoStop}%
\bibitem [{\citenamefont {Rocca}, \citenamefont {Lu},\ and\ \citenamefont
  {Galli}(2010)}]{Rocca2010}%
  \BibitemOpen
  \bibfield  {author} {\bibinfo {author} {\bibfnamefont {D.}~\bibnamefont
  {Rocca}}, \bibinfo {author} {\bibfnamefont {D.}~\bibnamefont {Lu}}, \ and\
  \bibinfo {author} {\bibfnamefont {G.}~\bibnamefont {Galli}},\ }\href@noop {}
  {\bibfield  {journal} {\bibinfo  {journal} {J. Chem. Phys.}\ }\textbf
  {\bibinfo {volume} {133}},\ \bibinfo {pages} {164109} (\bibinfo {year}
  {2010})}\BibitemShut {NoStop}%
\bibitem [{\citenamefont {Blase}\ and\ \citenamefont
  {Attaccalite}(2011)}]{Blase2011}%
  \BibitemOpen
  \bibfield  {author} {\bibinfo {author} {\bibfnamefont {X.}~\bibnamefont
  {Blase}}\ and\ \bibinfo {author} {\bibfnamefont {C.}~\bibnamefont
  {Attaccalite}},\ }\href@noop {} {\bibfield  {journal} {\bibinfo  {journal}
  {Appl. Phys. Lett.}\ }\textbf {\bibinfo {volume} {99}},\ \bibinfo {pages}
  {171909} (\bibinfo {year} {2011})}\BibitemShut {NoStop}%
\bibitem [{\citenamefont {Faber}\ \emph {et~al.}(2012)\citenamefont {Faber},
  \citenamefont {Duchemin}, \citenamefont {Deutsch}, \citenamefont
  {Attaccalite}, \citenamefont {Olevano},\ and\ \citenamefont
  {Blase}}]{Faber2012}%
  \BibitemOpen
  \bibfield  {author} {\bibinfo {author} {\bibfnamefont {C.}~\bibnamefont
  {Faber}}, \bibinfo {author} {\bibfnamefont {I.}~\bibnamefont {Duchemin}},
  \bibinfo {author} {\bibfnamefont {T.}~\bibnamefont {Deutsch}}, \bibinfo
  {author} {\bibfnamefont {C.}~\bibnamefont {Attaccalite}}, \bibinfo {author}
  {\bibfnamefont {V.}~\bibnamefont {Olevano}}, \ and\ \bibinfo {author}
  {\bibfnamefont {X.}~\bibnamefont {Blase}},\ }\href@noop {} {\bibfield
  {journal} {\bibinfo  {journal} {J. Mater. Sci.}\ }\textbf {\bibinfo {volume}
  {47}},\ \bibinfo {pages} {7472} (\bibinfo {year} {2012})}\BibitemShut
  {NoStop}%
\bibitem [{\citenamefont {Faber}\ \emph {et~al.}(2014)\citenamefont {Faber},
  \citenamefont {Boulanger}, \citenamefont {Attaccalite}, \citenamefont
  {Duchemin},\ and\ \citenamefont {Blase}}]{Faber2014}%
  \BibitemOpen
  \bibfield  {author} {\bibinfo {author} {\bibfnamefont {C.}~\bibnamefont
  {Faber}}, \bibinfo {author} {\bibfnamefont {P.}~\bibnamefont {Boulanger}},
  \bibinfo {author} {\bibfnamefont {C.}~\bibnamefont {Attaccalite}}, \bibinfo
  {author} {\bibfnamefont {I.}~\bibnamefont {Duchemin}}, \ and\ \bibinfo
  {author} {\bibfnamefont {X.}~\bibnamefont {Blase}},\ }\href@noop {}
  {\bibfield  {journal} {\bibinfo  {journal} {Philos. Trans. A Math. Phys. Eng.
  Sci.}\ }\textbf {\bibinfo {volume} {372}},\ \bibinfo {pages} {20130271}
  (\bibinfo {year} {2014})}\BibitemShut {NoStop}%
\bibitem [{\citenamefont {Walker}\ \emph {et~al.}(2006)\citenamefont {Walker},
  \citenamefont {Saitta}, \citenamefont {Gebauer},\ and\ \citenamefont
  {Baroni}}]{Walker2006}%
  \BibitemOpen
  \bibfield  {author} {\bibinfo {author} {\bibfnamefont {B.}~\bibnamefont
  {Walker}}, \bibinfo {author} {\bibfnamefont {A.~M.}\ \bibnamefont {Saitta}},
  \bibinfo {author} {\bibfnamefont {R.}~\bibnamefont {Gebauer}}, \ and\
  \bibinfo {author} {\bibfnamefont {S.}~\bibnamefont {Baroni}},\ }\href
  {http://journals.aps.org/prl/abstract/10.1103/PhysRevLett.96.113001}
  {\bibfield  {journal} {\bibinfo  {journal} {Phys. Rev. Lett.}\ }\textbf
  {\bibinfo {volume} {96}},\ \bibinfo {pages} {113001} (\bibinfo {year}
  {2006})}\BibitemShut {NoStop}%
\bibitem [{\citenamefont {Rocca}\ \emph {et~al.}(2008)\citenamefont {Rocca},
  \citenamefont {Gebauer}, \citenamefont {Saad},\ and\ \citenamefont
  {Baroni}}]{Rocca2008}%
  \BibitemOpen
  \bibfield  {author} {\bibinfo {author} {\bibfnamefont {D.}~\bibnamefont
  {Rocca}}, \bibinfo {author} {\bibfnamefont {R.}~\bibnamefont {Gebauer}},
  \bibinfo {author} {\bibfnamefont {Y.}~\bibnamefont {Saad}}, \ and\ \bibinfo
  {author} {\bibfnamefont {S.}~\bibnamefont {Baroni}},\ }\href@noop {}
  {\bibfield  {journal} {\bibinfo  {journal} {J. Chem. Phys.}\ }\textbf
  {\bibinfo {volume} {128}},\ \bibinfo {pages} {154105} (\bibinfo {year}
  {2008})}\BibitemShut {NoStop}%
\bibitem [{\citenamefont {Wilson}, \citenamefont {Gygi},\ and\ \citenamefont
  {Galli}(2008)}]{Wilson2008}%
  \BibitemOpen
  \bibfield  {author} {\bibinfo {author} {\bibfnamefont {H.~F.}\ \bibnamefont
  {Wilson}}, \bibinfo {author} {\bibfnamefont {F.}~\bibnamefont {Gygi}}, \ and\
  \bibinfo {author} {\bibfnamefont {G.}~\bibnamefont {Galli}},\ }\href
  {http://journals.aps.org/prb/abstract/10.1103/PhysRevB.78.113303} {\bibfield
  {journal} {\bibinfo  {journal} {Phys. Rev. B}\ }\textbf {\bibinfo {volume}
  {78}},\ \bibinfo {pages} {113303} (\bibinfo {year} {2008})}\BibitemShut
  {NoStop}%
\bibitem [{\citenamefont {Baroni}\ \emph {et~al.}(2001)\citenamefont {Baroni},
  \citenamefont {de~Gironcoli}, \citenamefont {Dal~Corso},\ and\ \citenamefont
  {Giannozzi}}]{Baroni2001}%
  \BibitemOpen
  \bibfield  {author} {\bibinfo {author} {\bibfnamefont {S.}~\bibnamefont
  {Baroni}}, \bibinfo {author} {\bibfnamefont {S.}~\bibnamefont
  {de~Gironcoli}}, \bibinfo {author} {\bibfnamefont {A.}~\bibnamefont
  {Dal~Corso}}, \ and\ \bibinfo {author} {\bibfnamefont {P.}~\bibnamefont
  {Giannozzi}},\ }\href@noop {} {\bibfield  {journal} {\bibinfo  {journal}
  {Rev. Mod. Phys.}\ }\textbf {\bibinfo {volume} {73}},\ \bibinfo {pages} {515}
  (\bibinfo {year} {2001})}\BibitemShut {NoStop}%
\bibitem [{\citenamefont {Baer}, \citenamefont {Neuhauser},\ and\ \citenamefont
  {Rabani}(2013)}]{Baer2013}%
  \BibitemOpen
  \bibfield  {author} {\bibinfo {author} {\bibfnamefont {R.}~\bibnamefont
  {Baer}}, \bibinfo {author} {\bibfnamefont {D.}~\bibnamefont {Neuhauser}}, \
  and\ \bibinfo {author} {\bibfnamefont {E.}~\bibnamefont {Rabani}},\ }\href
  {\doibase 10.1103/PhysRevLett.111.106402} {\bibfield  {journal} {\bibinfo
  {journal} {Phys. Rev. Lett.}\ }\textbf {\bibinfo {volume} {111}},\ \bibinfo
  {pages} {106402} (\bibinfo {year} {2013})}\BibitemShut {NoStop}%
\bibitem [{\citenamefont {Neuhauser}, \citenamefont {Baer},\ and\ \citenamefont
  {Rabani}(2014)}]{Neuhauser2014}%
  \BibitemOpen
  \bibfield  {author} {\bibinfo {author} {\bibfnamefont {D.}~\bibnamefont
  {Neuhauser}}, \bibinfo {author} {\bibfnamefont {R.}~\bibnamefont {Baer}}, \
  and\ \bibinfo {author} {\bibfnamefont {E.}~\bibnamefont {Rabani}},\
  }\href@noop {} {\bibfield  {journal} {\bibinfo  {journal} {J. Chem. Phys.}\
  }\textbf {\bibinfo {volume} {141}},\ \bibinfo {pages} {041102} (\bibinfo
  {year} {2014})}\BibitemShut {NoStop}%
\bibitem [{\citenamefont {Neuhauser}, \citenamefont {Rabani},\ and\
  \citenamefont {Baer}(2013{\natexlab{a}})}]{Neuhauser2013}%
  \BibitemOpen
  \bibfield  {author} {\bibinfo {author} {\bibfnamefont {D.}~\bibnamefont
  {Neuhauser}}, \bibinfo {author} {\bibfnamefont {E.}~\bibnamefont {Rabani}}, \
  and\ \bibinfo {author} {\bibfnamefont {R.}~\bibnamefont {Baer}},\ }\href@noop
  {} {\bibfield  {journal} {\bibinfo  {journal} {J. Chem. Theory Comput.}\
  }\textbf {\bibinfo {volume} {9}},\ \bibinfo {pages} {24} (\bibinfo {year}
  {2013}{\natexlab{a}})}\BibitemShut {NoStop}%
\bibitem [{\citenamefont {Ge}\ \emph {et~al.}(2013)\citenamefont {Ge},
  \citenamefont {Gao}, \citenamefont {Baer}, \citenamefont {Rabani},\ and\
  \citenamefont {Neuhauser}}]{Ge2013}%
  \BibitemOpen
  \bibfield  {author} {\bibinfo {author} {\bibfnamefont {Q.}~\bibnamefont
  {Ge}}, \bibinfo {author} {\bibfnamefont {Y.}~\bibnamefont {Gao}}, \bibinfo
  {author} {\bibfnamefont {R.}~\bibnamefont {Baer}}, \bibinfo {author}
  {\bibfnamefont {E.}~\bibnamefont {Rabani}}, \ and\ \bibinfo {author}
  {\bibfnamefont {D.}~\bibnamefont {Neuhauser}},\ }\href@noop {} {\bibfield
  {journal} {\bibinfo  {journal} {J. Phys. Chem. Lett.}\ }\textbf {\bibinfo
  {volume} {5}},\ \bibinfo {pages} {185} (\bibinfo {year} {2013})}\BibitemShut
  {NoStop}%
\bibitem [{\citenamefont {Neuhauser}, \citenamefont {Rabani},\ and\
  \citenamefont {Baer}(2013{\natexlab{b}})}]{Neuhauser2013a}%
  \BibitemOpen
  \bibfield  {author} {\bibinfo {author} {\bibfnamefont {D.}~\bibnamefont
  {Neuhauser}}, \bibinfo {author} {\bibfnamefont {E.}~\bibnamefont {Rabani}}, \
  and\ \bibinfo {author} {\bibfnamefont {R.}~\bibnamefont {Baer}},\ }\href@noop
  {} {\bibfield  {journal} {\bibinfo  {journal} {J. Phys. Chem. Lett.}\
  }\textbf {\bibinfo {volume} {4}},\ \bibinfo {pages} {1172} (\bibinfo {year}
  {2013}{\natexlab{b}})}\BibitemShut {NoStop}%
\bibitem [{\citenamefont {Baer}\ and\ \citenamefont
  {Rabani}(2012)}]{Baer2012a}%
  \BibitemOpen
  \bibfield  {author} {\bibinfo {author} {\bibfnamefont {R.}~\bibnamefont
  {Baer}}\ and\ \bibinfo {author} {\bibfnamefont {E.}~\bibnamefont {Rabani}},\
  }\href@noop {} {\bibfield  {journal} {\bibinfo  {journal} {Nano Lett.}\
  }\textbf {\bibinfo {volume} {12}},\ \bibinfo {pages} {2123} (\bibinfo {year}
  {2012})}\BibitemShut {NoStop}%
\bibitem [{\citenamefont {Neuhauser}\ \emph {et~al.}(2014)\citenamefont
  {Neuhauser}, \citenamefont {Gao}, \citenamefont {Arntsen}, \citenamefont
  {Karshenas}, \citenamefont {Rabani},\ and\ \citenamefont
  {Baer}}]{Neuhauser2014a}%
  \BibitemOpen
  \bibfield  {author} {\bibinfo {author} {\bibfnamefont {D.}~\bibnamefont
  {Neuhauser}}, \bibinfo {author} {\bibfnamefont {Y.}~\bibnamefont {Gao}},
  \bibinfo {author} {\bibfnamefont {C.}~\bibnamefont {Arntsen}}, \bibinfo
  {author} {\bibfnamefont {C.}~\bibnamefont {Karshenas}}, \bibinfo {author}
  {\bibfnamefont {E.}~\bibnamefont {Rabani}}, \ and\ \bibinfo {author}
  {\bibfnamefont {R.}~\bibnamefont {Baer}},\ }\href@noop {} {\bibfield
  {journal} {\bibinfo  {journal} {Phys. Rev. Lett.}\ }\textbf {\bibinfo
  {volume} {113}},\ \bibinfo {pages} {076402} (\bibinfo {year}
  {2014})}\BibitemShut {NoStop}%
\bibitem [{\citenamefont {Gao}\ \emph {et~al.}(2015)\citenamefont {Gao},
  \citenamefont {Neuhauser}, \citenamefont {Baer},\ and\ \citenamefont
  {Rabani}}]{Gao2014TDsDFT}%
  \BibitemOpen
  \bibfield  {author} {\bibinfo {author} {\bibfnamefont {Y.}~\bibnamefont
  {Gao}}, \bibinfo {author} {\bibfnamefont {D.}~\bibnamefont {Neuhauser}},
  \bibinfo {author} {\bibfnamefont {R.}~\bibnamefont {Baer}}, \ and\ \bibinfo
  {author} {\bibfnamefont {E.}~\bibnamefont {Rabani}},\ }\href@noop {}
  {\bibfield  {journal} {\bibinfo  {journal} {J. Chem. Phys.}\ }\textbf
  {\bibinfo {volume} {142}},\ \bibinfo {pages} {034106} (\bibinfo {year}
  {2015})}\BibitemShut {NoStop}%
\bibitem [{\citenamefont {Casida}(1995)}]{Casida1995}%
  \BibitemOpen
  \bibfield  {author} {\bibinfo {author} {\bibfnamefont {M.~E.}\ \bibnamefont
  {Casida}},\ }\href@noop {} {\bibfield  {journal} {\bibinfo  {journal} {Recent
  advances in density functional methods}\ }\textbf {\bibinfo {volume} {1}},\
  \bibinfo {pages} {155} (\bibinfo {year} {1995})}\BibitemShut {NoStop}%
\bibitem [{\citenamefont {Casida}(1996)}]{Casida1996}%
  \BibitemOpen
  \bibfield  {author} {\bibinfo {author} {\bibfnamefont {M.~E.}\ \bibnamefont
  {Casida}},\ }\enquote {\bibinfo {title} {Time-dependent density functional
  response theory of molecular systems: theory, computational methods, and
  functionals},}\ in\ \href@noop {} {\emph {\bibinfo {booktitle} {Recent
  Developments and Applications in Density Functional Theory}}},\ \bibinfo
  {editor} {edited by\ \bibinfo {editor} {\bibfnamefont {J.~M.}\ \bibnamefont
  {Seminario}}}\ (\bibinfo  {publisher} {Elsevier},\ \bibinfo {address}
  {Amsterdam},\ \bibinfo {year} {1996})\ pp.\ \bibinfo {pages}
  {391--439}\BibitemShut {NoStop}%
\bibitem [{\citenamefont {Furche}(2001)}]{Furche2001}%
  \BibitemOpen
  \bibfield  {author} {\bibinfo {author} {\bibfnamefont {F.}~\bibnamefont
  {Furche}},\ }\href@noop {} {\bibfield  {journal} {\bibinfo  {journal} {Phys.
  Rev. B}\ }\textbf {\bibinfo {volume} {64}},\ \bibinfo {pages} {195120}
  (\bibinfo {year} {2001})}\BibitemShut {NoStop}%
\bibitem [{\citenamefont {Dyson}(1953)}]{Dyson1953}%
  \BibitemOpen
  \bibfield  {author} {\bibinfo {author} {\bibfnamefont {F.~J.}\ \bibnamefont
  {Dyson}},\ }\href@noop {} {\bibfield  {journal} {\bibinfo  {journal} {Phys.
  Rev.}\ }\textbf {\bibinfo {volume} {90}},\ \bibinfo {pages} {994} (\bibinfo
  {year} {1953})}\BibitemShut {NoStop}%
\bibitem [{\citenamefont {Taylor}(1954)}]{Taylor1954}%
  \BibitemOpen
  \bibfield  {author} {\bibinfo {author} {\bibfnamefont {J.}~\bibnamefont
  {Taylor}},\ }\href@noop {} {\bibfield  {journal} {\bibinfo  {journal} {Phys.
  Rev.}\ }\textbf {\bibinfo {volume} {95}},\ \bibinfo {pages} {1313} (\bibinfo
  {year} {1954})}\BibitemShut {NoStop}%
\bibitem [{\citenamefont {Hirata}\ and\ \citenamefont
  {Head-Gordon}(1999{\natexlab{b}})}]{Hirata1999a}%
  \BibitemOpen
  \bibfield  {author} {\bibinfo {author} {\bibfnamefont {S.}~\bibnamefont
  {Hirata}}\ and\ \bibinfo {author} {\bibfnamefont {M.}~\bibnamefont
  {Head-Gordon}},\ }\href@noop {} {\bibfield  {journal} {\bibinfo  {journal}
  {Chem. Phys. Lett.}\ }\textbf {\bibinfo {volume} {314}},\ \bibinfo {pages}
  {291} (\bibinfo {year} {1999}{\natexlab{b}})}\BibitemShut {NoStop}%
\bibitem [{\citenamefont {Hirata}, \citenamefont {Head-Gordon},\ and\
  \citenamefont {Bartlett}(1999)}]{Hirata1999b}%
  \BibitemOpen
  \bibfield  {author} {\bibinfo {author} {\bibfnamefont {S.}~\bibnamefont
  {Hirata}}, \bibinfo {author} {\bibfnamefont {M.}~\bibnamefont {Head-Gordon}},
  \ and\ \bibinfo {author} {\bibfnamefont {R.~J.}\ \bibnamefont {Bartlett}},\
  }\href@noop {} {\bibfield  {journal} {\bibinfo  {journal} {J. Chem. Phys.}\
  }\textbf {\bibinfo {volume} {111}},\ \bibinfo {pages} {10774} (\bibinfo
  {year} {1999})}\BibitemShut {NoStop}%
\bibitem [{\citenamefont {Baer}\ and\ \citenamefont
  {Neuhauser}(2005)}]{Baer2005a}%
  \BibitemOpen
  \bibfield  {author} {\bibinfo {author} {\bibfnamefont {R.}~\bibnamefont
  {Baer}}\ and\ \bibinfo {author} {\bibfnamefont {D.}~\bibnamefont
  {Neuhauser}},\ }\href@noop {} {\bibfield  {journal} {\bibinfo  {journal}
  {Phys. Rev. Lett.}\ }\textbf {\bibinfo {volume} {94}},\ \bibinfo {pages}
  {043002} (\bibinfo {year} {2005})}\BibitemShut {NoStop}%
\bibitem [{\citenamefont {Brothers}\ \emph {et~al.}(2008)\citenamefont
  {Brothers}, \citenamefont {Izmaylov}, \citenamefont {Normand}, \citenamefont
  {Barone},\ and\ \citenamefont {Scuseria}}]{Brothers2008}%
  \BibitemOpen
  \bibfield  {author} {\bibinfo {author} {\bibfnamefont {E.~N.}\ \bibnamefont
  {Brothers}}, \bibinfo {author} {\bibfnamefont {A.~F.}\ \bibnamefont
  {Izmaylov}}, \bibinfo {author} {\bibfnamefont {J.~O.}\ \bibnamefont
  {Normand}}, \bibinfo {author} {\bibfnamefont {V.}~\bibnamefont {Barone}}, \
  and\ \bibinfo {author} {\bibfnamefont {G.~E.}\ \bibnamefont {Scuseria}},\
  }\href@noop {} {\bibfield  {journal} {\bibinfo  {journal} {J. Chem. Phys.}\
  }\textbf {\bibinfo {volume} {129}},\ \bibinfo {pages} {011102} (\bibinfo
  {year} {2008})}\BibitemShut {NoStop}%
\bibitem [{\citenamefont {Wang}\ and\ \citenamefont
  {Zunger}(1994{\natexlab{a}})}]{Wang1994d}%
  \BibitemOpen
  \bibfield  {author} {\bibinfo {author} {\bibfnamefont {L.~W.}\ \bibnamefont
  {Wang}}\ and\ \bibinfo {author} {\bibfnamefont {A.}~\bibnamefont {Zunger}},\
  }\href@noop {} {\bibfield  {journal} {\bibinfo  {journal} {J. Phys. Chem.}\
  }\textbf {\bibinfo {volume} {98}},\ \bibinfo {pages} {2158} (\bibinfo {year}
  {1994}{\natexlab{a}})}\BibitemShut {NoStop}%
\bibitem [{\citenamefont {Wang}\ and\ \citenamefont {Zunger}(1995)}]{Wang1995}%
  \BibitemOpen
  \bibfield  {author} {\bibinfo {author} {\bibfnamefont {L.~W.}\ \bibnamefont
  {Wang}}\ and\ \bibinfo {author} {\bibfnamefont {A.}~\bibnamefont {Zunger}},\
  }\href@noop {} {\bibfield  {journal} {\bibinfo  {journal} {Phys. Rev. B}\
  }\textbf {\bibinfo {volume} {51}},\ \bibinfo {pages} {17398} (\bibinfo {year}
  {1995})}\BibitemShut {NoStop}%
\bibitem [{\citenamefont {Wang}\ and\ \citenamefont {Zunger}(1996)}]{Wang1996}%
  \BibitemOpen
  \bibfield  {author} {\bibinfo {author} {\bibfnamefont {L.~W.}\ \bibnamefont
  {Wang}}\ and\ \bibinfo {author} {\bibfnamefont {A.}~\bibnamefont {Zunger}},\
  }\href@noop {} {\bibfield  {journal} {\bibinfo  {journal} {Phys. Rev. B}\
  }\textbf {\bibinfo {volume} {53}},\ \bibinfo {pages} {9579} (\bibinfo {year}
  {1996})}\BibitemShut {NoStop}%
\bibitem [{\citenamefont {Fu}\ and\ \citenamefont
  {Zunger}(1997{\natexlab{a}})}]{Fu1997}%
  \BibitemOpen
  \bibfield  {author} {\bibinfo {author} {\bibfnamefont {H.}~\bibnamefont
  {Fu}}\ and\ \bibinfo {author} {\bibfnamefont {A.}~\bibnamefont {Zunger}},\
  }\href@noop {} {\bibfield  {journal} {\bibinfo  {journal} {Phys. Rev. B}\
  }\textbf {\bibinfo {volume} {55}},\ \bibinfo {pages} {1642} (\bibinfo {year}
  {1997}{\natexlab{a}})}\BibitemShut {NoStop}%
\bibitem [{\citenamefont {Williamson}\ and\ \citenamefont
  {Zunger}(2000)}]{Williamson2000}%
  \BibitemOpen
  \bibfield  {author} {\bibinfo {author} {\bibfnamefont {A.~J.}\ \bibnamefont
  {Williamson}}\ and\ \bibinfo {author} {\bibfnamefont {A.}~\bibnamefont
  {Zunger}},\ }\href@noop {} {\bibfield  {journal} {\bibinfo  {journal} {Phys.
  Rev. B}\ }\textbf {\bibinfo {volume} {61}},\ \bibinfo {pages} {1978}
  (\bibinfo {year} {2000})}\BibitemShut {NoStop}%
\bibitem [{\citenamefont {Franceschetti}\ and\ \citenamefont
  {Zunger}(2000{\natexlab{a}})}]{Franceschetti2000a}%
  \BibitemOpen
  \bibfield  {author} {\bibinfo {author} {\bibfnamefont {A.}~\bibnamefont
  {Franceschetti}}\ and\ \bibinfo {author} {\bibfnamefont {A.}~\bibnamefont
  {Zunger}},\ }\href@noop {} {\bibfield  {journal} {\bibinfo  {journal} {Phys.
  Rev. B}\ }\textbf {\bibinfo {volume} {62}},\ \bibinfo {pages} {2614}
  (\bibinfo {year} {2000}{\natexlab{a}})}\BibitemShut {NoStop}%
\bibitem [{\citenamefont {Zunger}(2001)}]{Zunger2001}%
  \BibitemOpen
  \bibfield  {author} {\bibinfo {author} {\bibfnamefont {A.}~\bibnamefont
  {Zunger}},\ }\href@noop {} {\bibfield  {journal} {\bibinfo  {journal}
  {Physica Status Solidi B-Basic Research}\ }\textbf {\bibinfo {volume}
  {224}},\ \bibinfo {pages} {727} (\bibinfo {year} {2001})}\BibitemShut
  {NoStop}%
\bibitem [{\citenamefont {Fu}\ and\ \citenamefont
  {Zunger}(1997{\natexlab{b}})}]{Fu1997b}%
  \BibitemOpen
  \bibfield  {author} {\bibinfo {author} {\bibfnamefont {H.~X.}\ \bibnamefont
  {Fu}}\ and\ \bibinfo {author} {\bibfnamefont {A.}~\bibnamefont {Zunger}},\
  }\href@noop {} {\bibfield  {journal} {\bibinfo  {journal} {Phys. Rev. B}\
  }\textbf {\bibinfo {volume} {56}},\ \bibinfo {pages} {1496} (\bibinfo {year}
  {1997}{\natexlab{b}})}\BibitemShut {NoStop}%
\bibitem [{\citenamefont {Rabani}\ \emph {et~al.}(1999)\citenamefont {Rabani},
  \citenamefont {Hetenyi}, \citenamefont {Berne},\ and\ \citenamefont
  {Brus}}]{Rabani1999b}%
  \BibitemOpen
  \bibfield  {author} {\bibinfo {author} {\bibfnamefont {E.}~\bibnamefont
  {Rabani}}, \bibinfo {author} {\bibfnamefont {B.}~\bibnamefont {Hetenyi}},
  \bibinfo {author} {\bibfnamefont {B.~J.}\ \bibnamefont {Berne}}, \ and\
  \bibinfo {author} {\bibfnamefont {L.~E.}\ \bibnamefont {Brus}},\ }\href@noop
  {} {\bibfield  {journal} {\bibinfo  {journal} {J. Chem. Phys.}\ }\textbf
  {\bibinfo {volume} {110}},\ \bibinfo {pages} {5355} (\bibinfo {year}
  {1999})}\BibitemShut {NoStop}%
\bibitem [{\citenamefont {Reboredo}, \citenamefont {Franceschetti},\ and\
  \citenamefont {Zunger}(2000)}]{Reboredo2000}%
  \BibitemOpen
  \bibfield  {author} {\bibinfo {author} {\bibfnamefont {F.~A.}\ \bibnamefont
  {Reboredo}}, \bibinfo {author} {\bibfnamefont {A.}~\bibnamefont
  {Franceschetti}}, \ and\ \bibinfo {author} {\bibfnamefont {A.}~\bibnamefont
  {Zunger}},\ }\href@noop {} {\bibfield  {journal} {\bibinfo  {journal} {Phys.
  Rev. B}\ }\textbf {\bibinfo {volume} {61}},\ \bibinfo {pages} {13073}
  (\bibinfo {year} {2000})}\BibitemShut {NoStop}%
\bibitem [{\citenamefont {Franceschetti}\ and\ \citenamefont
  {Zunger}(2000{\natexlab{b}})}]{Franceschetti2000}%
  \BibitemOpen
  \bibfield  {author} {\bibinfo {author} {\bibfnamefont {A.}~\bibnamefont
  {Franceschetti}}\ and\ \bibinfo {author} {\bibfnamefont {A.}~\bibnamefont
  {Zunger}},\ }\href@noop {} {\bibfield  {journal} {\bibinfo  {journal} {Phys.
  Rev. B}\ }\textbf {\bibinfo {volume} {62}},\ \bibinfo {pages} {R16287}
  (\bibinfo {year} {2000}{\natexlab{b}})}\BibitemShut {NoStop}%
\bibitem [{\citenamefont {Eshet}, \citenamefont {Gr{\"u}nwald},\ and\
  \citenamefont {Rabani}(2013)}]{Eshet2013}%
  \BibitemOpen
  \bibfield  {author} {\bibinfo {author} {\bibfnamefont {H.}~\bibnamefont
  {Eshet}}, \bibinfo {author} {\bibfnamefont {M.}~\bibnamefont {Gr{\"u}nwald}},
  \ and\ \bibinfo {author} {\bibfnamefont {E.}~\bibnamefont {Rabani}},\
  }\href@noop {} {\bibfield  {journal} {\bibinfo  {journal} {Nano Lett.}\
  }\textbf {\bibinfo {volume} {13}},\ \bibinfo {pages} {5880} (\bibinfo {year}
  {2013})}\BibitemShut {NoStop}%
\bibitem [{\citenamefont {Wang}\ and\ \citenamefont
  {Zunger}(1994{\natexlab{b}})}]{Wang1994c}%
  \BibitemOpen
  \bibfield  {author} {\bibinfo {author} {\bibfnamefont {L.~W.}\ \bibnamefont
  {Wang}}\ and\ \bibinfo {author} {\bibfnamefont {A.}~\bibnamefont {Zunger}},\
  }\href@noop {} {\bibfield  {journal} {\bibinfo  {journal} {Phys. Rev. Lett.}\
  }\textbf {\bibinfo {volume} {73}},\ \bibinfo {pages} {1039} (\bibinfo {year}
  {1994}{\natexlab{b}})}\BibitemShut {NoStop}%
\bibitem [{\citenamefont {Zunger}\ and\ \citenamefont
  {Wang}(1996)}]{Zunger1996}%
  \BibitemOpen
  \bibfield  {author} {\bibinfo {author} {\bibfnamefont {A.}~\bibnamefont
  {Zunger}}\ and\ \bibinfo {author} {\bibfnamefont {L.~W.}\ \bibnamefont
  {Wang}},\ }\href@noop {} {\bibfield  {journal} {\bibinfo  {journal} {Appl.
  Surf. Sci.}\ }\textbf {\bibinfo {volume} {102}},\ \bibinfo {pages} {350}
  (\bibinfo {year} {1996})}\BibitemShut {NoStop}%
\bibitem [{\citenamefont {Kosloff}(1988)}]{Kosloff1988}%
  \BibitemOpen
  \bibfield  {author} {\bibinfo {author} {\bibfnamefont {R.}~\bibnamefont
  {Kosloff}},\ }\href@noop {} {\bibfield  {journal} {\bibinfo  {journal} {J.
  Phys. Chem.}\ }\textbf {\bibinfo {volume} {92}},\ \bibinfo {pages} {2087}
  (\bibinfo {year} {1988})}\BibitemShut {NoStop}%
\end{thebibliography}
\end{document}